\documentclass[twocolumn,pra,superscriptaddress]{revtex4-2}
\usepackage{amsmath,amssymb,mathrsfs}
\usepackage{natbib}
\usepackage{subfigure}
\usepackage{tabularx}
\usepackage{epsfig}
\usepackage{longtable}
\usepackage{amsfonts}
\usepackage{rotating}
\usepackage{bbold}
\usepackage{hhline}
\usepackage{braket}
\usepackage{txfonts, comment}
\usepackage{mathtools}

\usepackage[unicode=true,bookmarks=true,bookmarksnumbered=false,bookmarksopen=false,breaklinks=false,pdfborder={0 0 1},backref=false,colorlinks=true]{hyperref}

\hypersetup{linkcolor=magenta,urlcolor=blue,citecolor=blue,pdfstartview={FitH},hyperfootnotes=false,unicode=true}

\newtheorem{definition}{Definition}

\def\be{\begin{equation}}
\def\ee{\end{equation}}
\def\bea{\begin{eqnarray}}
\def\eea{\end{eqnarray}}

\usepackage[normalem]{ulem}

\begin{document}
\title{Quantum complexity phase transition in fermionic quantum circuits}

\author{Wei Xia}
\affiliation{Department of Physics and The Hong Kong Institute of Quantum Information Science and Technology,
The Chinese University of Hong Kong, Shatin, New Territories, Hong Kong, China}

\author{Yijia Zhou}
\affiliation{Shanghai Qi Zhi Institute, AI Tower, Xuhui District, Shanghai 200232, China}

\author{Xingze Qiu}
\email{xingze@tongji.edu.cn}
\affiliation{School of Physics Science and Engineering, Tongji University, Shanghai 200092, China}

\author{Xiaopeng Li}
\email{xiaopeng\_li@fudan.edu.cn}
\affiliation{Shanghai Qi Zhi Institute, AI Tower, Xuhui District, Shanghai 200232, China}
\affiliation{State Key Laboratory of Surface Physics, Institute of Nanoelectronics and Quantum Computing,
and Department of Physics, Fudan University, Shanghai 200433, China
}
\affiliation{Shanghai Artificial Intelligence Laboratory, Shanghai 200232, China} 
\affiliation{Shanghai Research Center for Quantum Sciences, Shanghai 201315, China} 
\affiliation{Hefei National Laboratory, Hefei 230088, China}

\begin{abstract} 
Understanding the complexity of quantum many-body systems has been attracting much attention recently for its fundamental importance in characterizing complex quantum phases beyond the scope of quantum entanglement. 
Here, we investigate Krylov complexity in quantum percolation models (QPM) and establish unconventional phase transitions emergent from the interplay of exponential scaling of the Krylov complexity and the number of spanning clusters in QPM. We develop a general scaling theory for Krylov complexity phase transitions  (KCPT) on QPM, and obtain exact results for the critical probabilities and exponents. For non-interacting systems across diverse lattices (1D/2D/3D regular, Bethe, and quasicrystals), our scaling theory reveals that the KCPT coincides with the classical percolation transition. In contrast, for interacting systems, we find the KCPT develops a generic separation from the percolation transition due to the highly complex quantum many-body effects, which is analogous to the Griffiths effect in the critical disorder phase transition. To test our theoretical predictions, we provide a concrete protocol for measuring the Krylov complexity, which is accessible to present experiments.

\end{abstract}

\date{\today}
\maketitle

{\it Introduction.---}
The quantum complexity phase transition is a novel concept emerging at the boundary between quantum and classical computation, forming part of Aaronson’s ``Ten Semi-Grand Challenges for Quantum Computing Theory"~\cite{aaronson2005Ten}. 
For instance, while Clifford circuits are classically simulable~\cite{Aaronson2004Improved,gottesman1998heisenberg,nielsen2010quantum}, adding $T$ gates induces universality and computational hardness, marking a transition characterized by quantum magic~\cite{Liu2022Many,Heinrich2019robustnessofmagic,Liu2019OneShot,Wang2020Efficiently,Heimendahl2021stabilizerextentis,Veitch2014resource,seddon2019quantifying}. Similarly, single-qubit measurements on $k$-regular graph states exhibit a sharp shift from efficient simulation to hardness at $k=3$, governed by entanglement width~\cite{Ghosh2015Complexity}. Recently, the monitored random circuits have attracted a lot of attention. Several types of complexity phase transitions are established in this model, including entanglement phase transitions~\cite{Skinner_2019,koh2023measurement,Choi_2020, Gullans_2020}, magic phase transitions~\cite{Fux2024Entanglement}, and C-complexity phase transitions~\cite{suzuki2025quantum}. 
While complexity exhibits intricate manifestations at phase transition criticality~\cite{niroula2024phase}, it remains an open question whether quantum complexity can serve as a driving force to distinguish otherwise indistinguishable many-body phases.

\begin{figure}
    \centering
\includegraphics[width=1\linewidth]{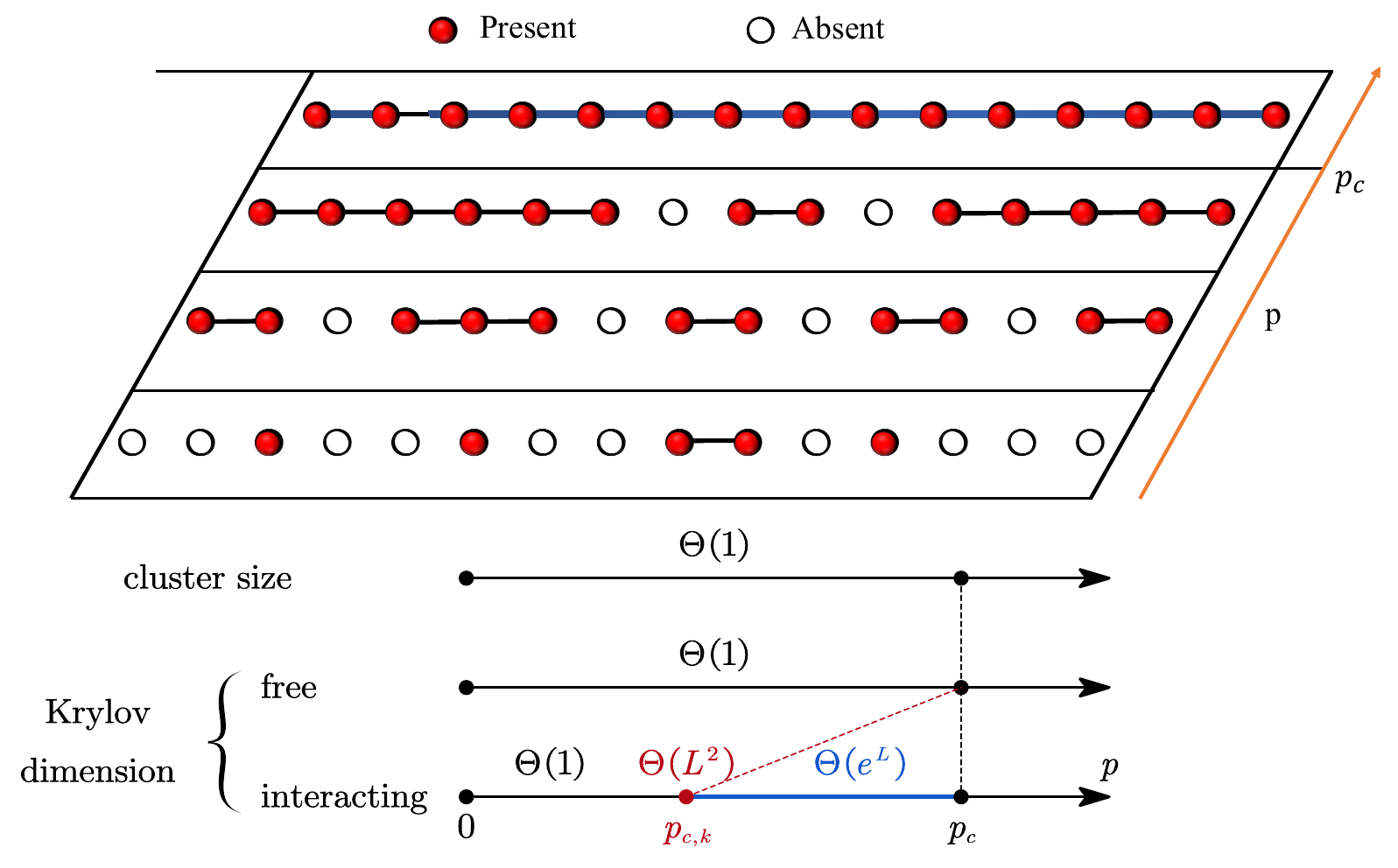}
    \caption{
    (a) Schematic of a percolation lattice. Sites are occupied (red) with probability $p$ or unoccupied (empty). Contiguous occupied sites form clusters; a spanning cluster (blue paths) connects system boundaries. (b) Phase diagram for QPMs on a 1D regular lattice. The classical critical point $p_c$ marks spanning cluster formation. For free fermions, Krylov dimension diverges at $p_c$. For interacting fermions, a distinct transition occurs at $p_{c,k}$: Krylov dimension is constant for $p < p_{c,k}$, grows quadratically at $p_{c,k}$, and diverges exponentially for $p > p_{c,k}$. 
   }
    \label{fig:PhaseTransition}
\end{figure}

Beyond the challenges of simulation, the complexity of general quantum dynamics is closely related to the concept of quantum information scrambling, which is commonly characterized by the out-of-time-order correlator (OTOC)~\cite{hashimoto2017out,swingle2018unscrambling,Li2017Measuring,Green2022Experimental,lewis2019dynamics,landsman2019verified,Xiao2021Information}. However, OTOCs lack exponential growth outside semiclassical limits, rendering Lyapunov exponents ill-defined for generic quantum systems~\cite{Parker2019Universal,bhattacharjee2022krylov}. Krylov complexity has been applied to various phase transitions, especially dynamical ones~\cite{rabinovici2021operator,rabinovici2022krylov,hashimoto2023krylov,Fabian2022Krylov,avdoshkin2022krylov,kundu2023state,bhattacharyya2023operator,caputa2024krylov,Bento2024Krylov}. 
Although Krylov complexity has been used to study various known phase transitions, a phase transition driven by Krylov complexity itself has not yet been established. Due to its computational difficulty, most studies rely on numerical simulations, with only a few models allowing analytical treatment~\cite{Parker2019Universal}. Experimental measurement remains challenging, as the complex structure of the Krylov basis hinders practical implementation. So far, no experimental protocols for measuring Krylov complexity have been proposed.

In this work, we investigate QPMs (Fig.\ref{fig:PhaseTransition}), which allow both analytical and numerical study of Krylov complexity. Within these models, we identify a quantum phase transition driven by Krylov complexity. Sites are occupied with probability $p$ (red dots) or unoccupied (empty circles), forming clusters. 
A spanning cluster (blue paths) signals the classical percolation transition at a critical probability $p_c$ (Supplementary Material).
With this lattice, we define a fermionic Hamiltonian [Eq.~\eqref{eq:Hamiltonian}] and leverage percolation theory to show that the long-time averaged Krylov complexity undergoes a phase transition. For non-interacting systems, the classical percolation transition induces a corresponding Krylov complexity transition at $p=p_c$, and we develop a scaling theory for this quantum transition. With interactions, this correspondence breaks: the Krylov transition occurs at $p_{c,k} < p_c$ due to exponential complexity growth amplifying rare events. Finally, we propose an experimental protocol using near-term quantum platforms (superconducting qubits, Rydberg arrays, trapped ions) involving (1) constructing a set of Krylov basis states that are
experimentally accessible and (2) performing an orthogonalization procedure on this set to recover the Krylov basis. This work establishes quantum percolation models as a powerful analytical and experimental platform for discovering and characterizing Krylov complexity phase transitions. We move beyond merely analyzing known transitions in specific models, demonstrating that complexity itself can drive a new class of quantum phase transitions. Crucially, we bridge a critical gap by proposing the first experimental protocol for measuring Krylov complexity on near-term quantum devices.

{\it Setup.---}
We consider the complexity phase transition in the QPMs. 
QPMs include site and bond variants \cite{stauffer2018introduction,essam1980percolation,Shante01051971}, with site percolation being more general~\cite{Zhen2023Direct}. These models probe disorder-driven transitions in localization, superconductivity, and quantum transport~\cite{mookerjee1995quantum,Shapir1982Localization,Soukoulis1991Localization,Khmel1984Localization,Kirkpatrick1972Localized,Lee1993Quantum}. Here, we consider site percolation: each site is occupied with probability $p$ or vacant with probability $1-p$. The tight-binding Hamiltonian is~\cite{Shapir1982Localization}
\begin{equation}\label{eq:Hamiltonian}
\textstyle    H = \sum_{\langle i,j \rangle} t_{ij} a_i^\dag a_j + u_{ij} n_i n_j,
\end{equation}
where $a_i^\dag$ creates a fermion at site $i$, $n_i = a_i^\dag a_i$, and $\langle i,j \rangle$ denotes nearest neighbors. We set $t_{ij} = u_{ij}$ for simplicity, with $t_{ij} = 1$ if sites $i,j$ are occupied and $0$ otherwise. 
Increasing $p$ induces a classical percolation transition at $p_c$, where spanning clusters emerge. The cluster number density $\rho(s, p)$ defines the probability $s \rho(s, p)$ that a random site belongs to a size-$s$ cluster (Supplementary Material). To capture the dynamical features of QPMs, we use quantum Krylov complexity. Our main goal is to identify and characterize a new phase transition in Krylov complexity, driven by percolation and interactions.

For an operator $O$ evolving under Hamiltonian $H$, the Heisenberg dynamics are~\cite{Parker2019Universal}  
\begin{equation}
    O(t) = e^{i\mathcal{L}t} O = \sum_{m=0}^\infty \frac{(it)^m}{m!} \mathcal{L}^m |O),
\end{equation}
where $\mathcal{L} |O) = |[H,O])$ is the Liouvillian, and $|O) = \sum_{ij} O_{ij} \ket{i}\!\bra{j}$ is the fictitious operator state based on an orthonormal basis $\{\ket{i}\}$. The Krylov space, spanned by $\{\mathcal{L}^m|O)\}$, is orthogonalized via the Lanczos algorithm to yield the basis $\{|k_m)\}$, where $\mathcal{L}$ is tridiagonal with Lanczos coefficients $b_m$. Expanding $O(t)$ gives $ |O(t)) = \sum_{m = 0}^{D-1} \phi_m(t)|k_m)$ with $D$ the Krylov dimension. The coefficients obey $ -i\partial_t \phi_m(t) = b_m \phi_{m-1}(t) + b_{m+1} \phi_{m+1}(t)$, subject to $\phi_m(0) = \delta_{m0}$, $b_0 = 0$. 
The Krylov complexity is then defined as the average position of the propagating packet over the Krylov chain: 
\begin{equation}
    C(t) = \sum_{m=0}^{D-1} m |\phi_m(t)|^2.
\end{equation}
Krylov complexity has been widely studied in areas such as quantum chaos and integrable systems~\cite{rabinovici2021operator,rabinovici2022krylov,rabinovici2022krylov,hashimoto2023krylov,Fabian2022Krylov,Kim2022Operator}, quantum field theory~\cite{barbon2019evolution,Dymarsky2021Krylov,avdoshkin2022krylov,kundu2023state}, and open quantum systems~\cite{bhattacharya2022operator,bhattacharjee2023operator,bhattacharya2023krylov,bhattacharjee2024operator,Liu2023Krylov}. It has also been suggested as a potential order parameter for dynamical phase transitions. However, a phase transition uniquely defined by Krylov complexity has not yet been clearly identified. In this work, we propose and systematically explore the emergence of such a transition in quantum Krylov complexity.

{\it Complexity phase transition.---}
We outline the complexity phase transition within QPMs.
Consider a general lattice in the thermodynamic limit with a local operator $O = n_i$. In non-interacting systems below the percolation threshold $p_c$, the system breaks into isolated clusters. Since the Hamiltonians of these clusters commute, the Krylov complexity of $n_i$ depends only on the cluster containing the initial operator. In the long-time limit, the wave packet spreads over the entire Krylov chain, making the complexity proportional to the Krylov dimension~\cite{xia2024complexity}. For $p < p_c$, cluster sizes are finite, so Krylov complexity remains finite.
When $p > p_c$, spanning clusters form, and their Krylov dimensions diverge. As a result, Krylov complexity also diverges in the long-time limit. This indicates that the classical percolation transition induces a phase transition in Krylov complexity for non-interacting systems. In the interacting case, however, the critical probability for the Krylov transition, $p_{c,k}$, is lower than $p_c$. This shift occurs because interactions change the scaling of Krylov complexity with system size from polynomial to exponential.

To quantify the divergence and the shift in the critical point, we examine the critical behavior of the long-time averaged Krylov complexity $\overline{C}$, which is determined by the scaling behavior and the distribution of cluster sizes. Specifically, $\overline{C} \propto D$, where $D$ is the Krylov dimension of the cluster containing the operator $n_i$. For non-interacting systems, $D \propto s^2$; for interacting systems, $D \propto 4^s$. Averaging over clusters:  
\begin{equation} \label{eq:KryDim}
    \overline{C} \propto D = \lim_{L\to\infty} \sum_{s=1}^{L} [s \rho(s, p)]  d(s) \sim \frac{\mathrm{poly}(p)}{|p - p_{c,k}|^{\delta}},
\end{equation}
where $d(s)$ is the cluster-size contribution, $s \rho(s, p)$ is the probability that $n_i$ resides in a size-$s$ cluster. 
The summation averages over all possible initial operator locations within a system of size $L$ (much greater than the mean cluster size). 
Taking $L\to\infty$, $ D $ has the generalized form $ {\rm poly}(p)/|p - p_{c,k}|^\delta $ near the critical point, where $  {\rm poly}(p) $ depends on $ d(s) $ and the lattice geometry, and the critical point $p_{c,k}$ marks divergence in $D$. The interplay between percolation and interactions plays a central role in shaping complexity dynamics. To explore this, we study two key cases on the percolation lattice: free fermions, where complexity is determined solely by the percolation transition, and interacting fermions, where many-body effects compete with percolation, potentially giving rise to more intricate complexity phase behavior.

\begin{figure}
    \centering
    \includegraphics[width=1\linewidth]{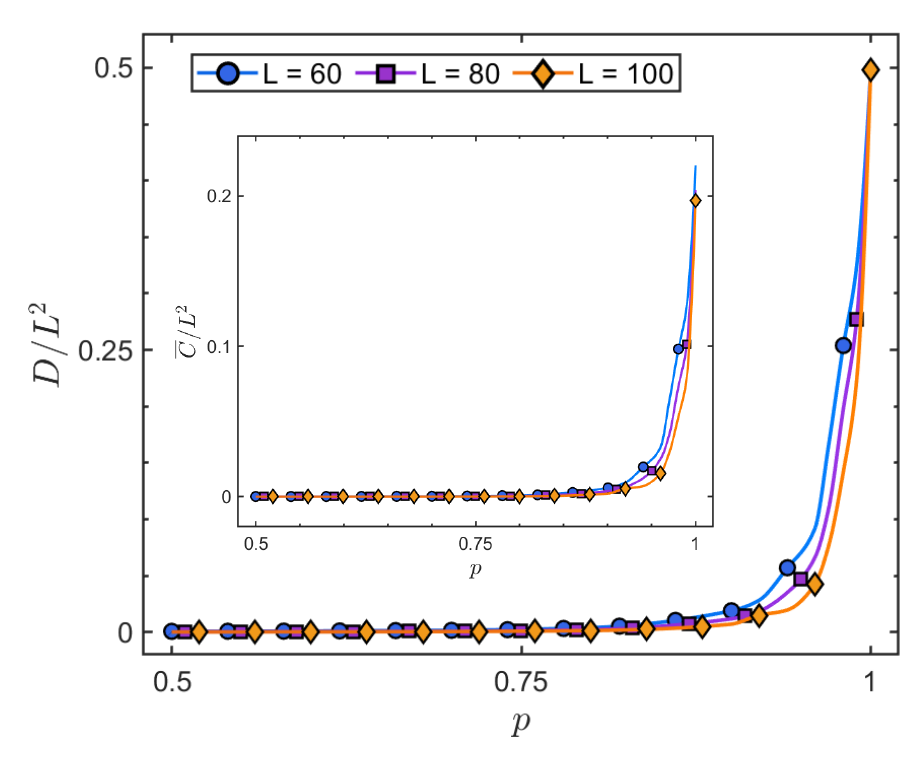}
    \caption{Krylov dimension $D$ and complexity $\overline{C}$ for 1D QPMs (the system size $L=60,80,100$). Results averaged over 200 configurations and initial operators per $p$, with time evolution to $t=10^5$ until saturation. We have averaged the evolution time when Krylov complexity reaches a plateau. }
    \label{fig:OneDimension}
\end{figure}

{\it Free fermion case.---}
For free fermions, $D \propto s^2$. At $p_c$, Krylov complexity diverges due to percolation. Using the cluster density scaling form~\cite{malthe2024percolation}  
\begin{equation}
    \rho(s, p) = s^{-\tau} \hat{F}((p_c - p) s^{\sigma}),
\end{equation}
where $\hat{F}(u)$ exponentially decaying, and $\tau$ and $\sigma$ are critical exponents. 
Substituting into Eq.~\eqref{eq:KryDim} yields  
\begin{equation}\label{eq:Scaling}
    D \sim \sum_{s=1}^{\infty} s^{3-\tau} \hat{F}((p_c - p) s^{\sigma}) \sim |p - p_{c,k}|^\delta,
\end{equation}
where $p_{c,k} = p_c$ and $\delta = (4-\tau)/\sigma$. 
This holds for general lattice geometries, including regular lattices, Bethe lattices, and quasicrystals (Supplementary Material).

As an example, we here consider the one-dimensional regular lattice. 
In this case, the critical exponents $\tau=2$ and $\sigma=1$, and the clusters of size $s$ require $s$ occupied sites bounded by vacancies, giving 
\begin{equation}
    \rho(s, p) = (1-p)^2 p^s.
\end{equation}
The contribution from a cluster of size $s$ is $d(s) \propto s^2$ and we have 
\begin{equation}
    D^{\text{Free}}_{\text{1D}} \propto \frac{p (p^2 + 4p + 1)}{(p - 1)^2},
\end{equation}
diverging at $p_{c,k} = 1$ with $\delta = 2$, consistent with $\delta = (4-\tau)/\sigma$. 

To verify this behavior numerically, we simulate a one-dimensional lattice of size up to $L = 100$. Fig.~\ref{fig:OneDimension}  shows the expectation values of the local operator $O_i = a_i^\dagger a_i = n_i$, averaged over different disorder configurations and site indices. For each value of $p$, we average over both disorder realizations and initial operator positions. The results show a sharp increase in both Krylov complexity and Krylov dimension as $p \to 1$, in agreement with theoretical predictions.

{\it Interacting fermion case.---}
The introduction of interactions breaks the correspondence between the Krylov complexity phase transition and the classical percolation transition. This effect is especially clear in the one-dimensional case, where interactions can cause the Krylov dimension to diverge exponentially. In 1D, with $d(s) \propto 4^s$, the Krylov dimension scales as

\begin{equation}
    D^{\text{Int}}_{1D} \propto
\begin{dcases}  
\frac{4p(1-p)^2}{(4p - 1)^2} & p < p_{c,k}, \\
L(L+1) & p = p_{c,k}, \\
L(4p)^{L+1} \frac{1 - p}{4p - 1} & p > p_{c,k},
\end{dcases}
\end{equation}
with a critical threshold $p_{c,k} = 1/4$, which differs from the classical percolation threshold $p_c = 1$. Since interactions are confined within each cluster, they do not affect the classical percolation transition at $p_c$. However, they significantly lower the critical point for the Krylov complexity transition, highlighting a key difference between the two transitions in the interacting case. In the classical regime ($p < p_{c,k}$), large clusters are extremely rare with exponentially small probability, and the Krylov dimension remains effectively bounded. In contrast, in the quantum regime ($p_{c,k} < p < p_c$), the average cluster size remains finite, but the Krylov dimension grows exponentially with system size.

This behavior arises because interactions enhance the impact of exponentially rare spanning clusters in the percolation process. For $p < p_c$, the probability of forming a spanning cluster decreases exponentially with system size. In non-interacting models, Krylov complexity grows only polynomially, and thus the exponentially suppressed probability does not significantly affect the critical behavior, leaving the critical probability unchanged. In contrast, in interacting models, Krylov complexity grows exponentially with system size. Therefore, even though spanning clusters are exponentially rare, their exponentially large contribution to the complexity becomes significant, effectively lowering the critical threshold.

\begin{figure*}
    \centering
    \includegraphics[width=.85\linewidth]{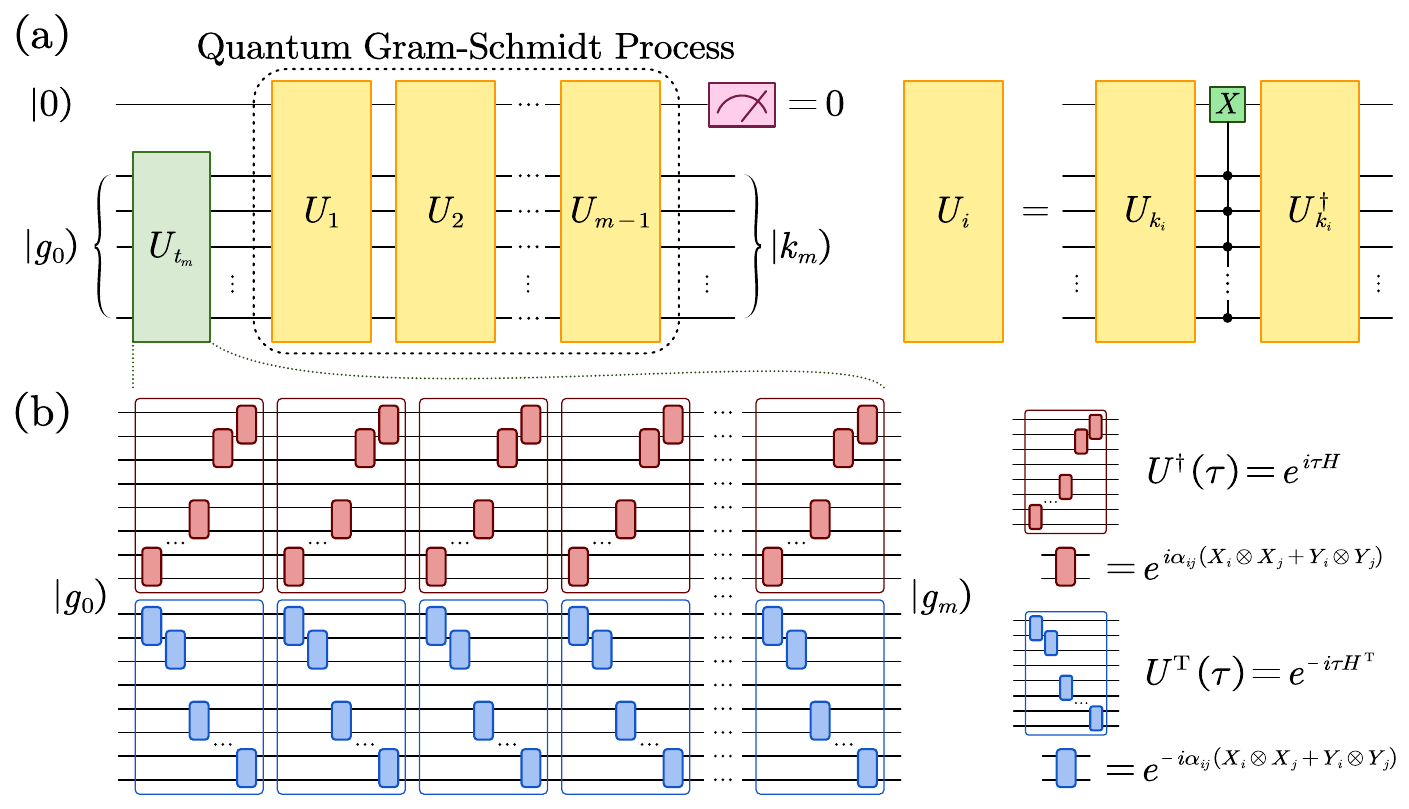}
    \caption{Schematics of our experimental protocol for measuring the Krylov complexity. (a) Quantum Gram-Schmidt circuit: Input $ |0) |g_0) $ outputs the orthogonalized Krylov basis $ |k_m) $. $U_{k_i}$ prepares the orthogonalized Krylov basis $ |k_i) $ and $ |g_m) $ is projected onto the orthogonalized operator state $ |k_i) $ by $ U_{i}$. (b) Preparing the operator state $ |g_m) $ with $U_{t_m} = \exp(i\mathcal{L}t_m)$. Here, we take the nearest-neighbor free fermionic models as an example. 
    }
    \label{fig:ExperimentProposal}
\end{figure*}

This behavior exhibits a profound connection to Griffiths effects in disordered systems~\cite{Griffiths1969Nonanalytic,vojta2010quantum}. Specifically, in the quantum regime $(p_{c,k} < p < p_c)$, exponentially rare spanning clusters – analogous to Griffiths regions – dominate the average Krylov dimension despite their vanishingly small probability. Interactions act as a crucial amplifier: within these rare, large clusters, the local Krylov complexity grows exponentially with cluster size $(d(s) \propto 4^s)$, while in non-interacting systems it grows only polynomially. This creates an exponential vs. exponential competition: the probability of finding a spanning cluster decays exponentially with system size $L(\sim e^{-cL})$, but its contribution to the global Krylov dimension $D^{\text{Int}}_{1D}$ grows exponentially $(\sim e^{kL})$. Crucially, for $p > p_{c,k}$, the exponential growth factor $k$ overcomes the exponential suppression factor $c$, meaning these rare Griffiths-like regions dictate the scaling of the average Krylov dimension. This avalanche effect, where interactions magnify the impact of exponentially rare fluctuations, is the fundamental reason for the drastic lowering of the Krylov critical threshold $(p_{c,k} = 1/4)$ compared to the classical percolation threshold $(p_c = 1)$. It highlights that interactions can induce a distinct quantum Griffiths phase characterized by exponentially diverging Krylov complexity driven by rare regions, decoupling the Krylov transition governed by rare-event statistics amplified by interactions from the underlying geometric (percolation) transition. The separation of critical points $(p_{c,k} < p_c)$ is a stark signature of how interactions fundamentally alter quantum critical dynamics beyond what classical percolation predicts.

{\it Experimental protocol.---}
Direct measurement of Krylov complexity is challenging, as there is no obvious map between the Krylov complexity and normal observables. 
Here, we propose a protocol for measuring Krylov complexity, which involves three steps. In the following discussion, we use the free fermion model as a representative example to illustrate our experimental proposal. However, the proposed scheme is general and can be applied to a broad class of models by modifying the Hamiltonian simulation subroutine.

Firstly, construct accessible states: Generate $|g_i) = \exp(i \mathcal{L} t_i)|g_0)$, where the time is discretized as $t_i = i \delta t,\, i = 0,1,\dots,D$ and $|g_0)$ is the initial state. These operator states have been shown to be linearly independent and are equivalent to the Krylov space~\cite{vcindrak2024measurable}.
To prepare these operator states on near-term quantum platforms, we need to map $e^{iHt} O e^{-iHt}$ to $e^{iHt} \otimes (e^{-iHt})^T \ket{O}$ in doubled Hilbert space. 
And for nearest-neighbor free fermions, the Jordan-Wigner transformation maps the system to a simple XY spin model, which is amenable to experimental implementation. To realize the time evolution, we employ a Trotter decomposition, where each Trotter step consists of quantum circuits built from the elementary gate $\exp(-i\alpha (X \otimes X + Y \otimes Y))$ shown in Fig.~\ref{fig:ExperimentProposal} (b).  

Secondly, Orthogonalize these operator states $|g_i)$: Apply quantum Gram-Schmidt [Fig.~\ref{fig:ExperimentProposal}(a)] using unitaries $U_{k_i}$ to prepare $|k_i)$ and $ U_{i}$ is applied to project $ |g_m) $ onto the orthogonalized operator state $ |k_i) $ (Supplementary Material). Note that preparing these orthogonalized operator states demands only polynomial resources~\cite{meng2020improved,Zhang2021Quantum,li2025quantumalgorithmvectorset}. 
At last, measure overlaps $|\phi_m(t)|^2$ between the measured states and the orthogonalized operator states to obtain the Krylov complexity. Our experimental approach provides a general method for measuring Krylov complexity and is compatible with superconducting qubits~\cite{huang2020superconducting,Krantz2019quantum,Kjaergaard2020Superconducting,Wendin2017Quantum}, Rydberg atom arrays~\cite{Wu2021concise,Adams2020Rydberg,Saffman2010Quantum,Saffman2016Quantumcomputing}, and trapped-ion systems~\cite{Haffner2008Quantum,Bruzewicz2019Trapped,blatt2012quantum,Monroe2021Programmable,Duan2010Colloquium,Monroe2013Scaling}.

{\it Conclusion.---}
We established a framework for Krylov complexity phase transitions in QPMs, showing that the long-time averaged Krylov complexity $\overline{C} \propto \sum_{s=1}^{\infty} [s n(s, p)]  d(s) \sim \mathrm{poly}(p)/|p - p_{c,k}|^{\delta}$. For free fermions on any lattice, $p_{c,k} = p_c$, indicating that classical percolation phase transitions can induce the corresponding Krylov complexity phase transition. In 1D interacting systems, $p_{c,k} = 1/4 < p_c$, as interactions reduce the threshold via altering the scaling behavior of Krylov complexity from polynomial to exponential. 
We designed an experimental protocol for near-term devices to detect the Krylov complexity and explore the corresponding complexity phase transitions.

{\it Acknowledgement.---}
Thank you for the effective discussion with Saud Čindrak, Adrian Paschke, Ziming Li, Ze Wu, Jie Zou and Chong Chen. This work is supported by National Program on Key Basic Research Project of China (Grant No. 2021YFA1400900), National Natural Science Foundation of China (Grant No.11774067, and 12304555), 
the Innovation Program for Quantum Science and Technology of China (Grant No. 2024ZD0300100),
Shanghai Municipal Science and Technology Major Project (Grant No. 25TQ003, 2019SHZDZX01, 24DP2600100). W.X. acknowledges support from the National Natural Science Foundation of China/Hong Kong RGC Collaborative Research Scheme (Project CRS CUHK401/22) and New Cornerstone
Science Foundation, Hong Kong RGC Senior Research Fellow Scheme, Ref. SRFS2223-4S01. Y.Z. acknowledges support from the Shanghai Qi Zhi Institute Innovation Program SQZ202317.


\begin{thebibliography}{77}%
\makeatletter
\providecommand \@ifxundefined [1]{%
 \@ifx{#1\undefined}
}%
\providecommand \@ifnum [1]{%
 \ifnum #1\expandafter \@firstoftwo
 \else \expandafter \@secondoftwo
 \fi
}%
\providecommand \@ifx [1]{%
 \ifx #1\expandafter \@firstoftwo
 \else \expandafter \@secondoftwo
 \fi
}%
\providecommand \natexlab [1]{#1}%
\providecommand \enquote  [1]{``#1''}%
\providecommand \bibnamefont  [1]{#1}%
\providecommand \bibfnamefont [1]{#1}%
\providecommand \citenamefont [1]{#1}%
\providecommand \href@noop [0]{\@secondoftwo}%
\providecommand \href [0]{\begingroup \@sanitize@url \@href}%
\providecommand \@href[1]{\@@startlink{#1}\@@href}%
\providecommand \@@href[1]{\endgroup#1\@@endlink}%
\providecommand \@sanitize@url [0]{\catcode `\\12\catcode `\$12\catcode
  `\&12\catcode `\#12\catcode `\^12\catcode `\_12\catcode `\%12\relax}%
\providecommand \@@startlink[1]{}%
\providecommand \@@endlink[0]{}%
\providecommand \url  [0]{\begingroup\@sanitize@url \@url }%
\providecommand \@url [1]{\endgroup\@href {#1}{\urlprefix }}%
\providecommand \urlprefix  [0]{URL }%
\providecommand \Eprint [0]{\href }%
\providecommand \doibase [0]{https://doi.org/}%
\providecommand \selectlanguage [0]{\@gobble}%
\providecommand \bibinfo  [0]{\@secondoftwo}%
\providecommand \bibfield  [0]{\@secondoftwo}%
\providecommand \translation [1]{[#1]}%
\providecommand \BibitemOpen [0]{}%
\providecommand \bibitemStop [0]{}%
\providecommand \bibitemNoStop [0]{.\EOS\space}%
\providecommand \EOS [0]{\spacefactor3000\relax}%
\providecommand \BibitemShut  [1]{\csname bibitem#1\endcsname}%
\let\auto@bib@innerbib\@empty
\bibitem [{\citenamefont {Aaronson}(2005)}]{aaronson2005Ten}%
  \BibitemOpen
  \bibfield  {author} {\bibinfo {author} {\bibfnamefont {S.}~\bibnamefont
  {Aaronson}},\ }\href {https://www.scottaaronson.com/writings/qchallenge.html}
  {\bibinfo {title} {Ten semi-grand challenges for quantum computing theory}}
  (\bibinfo {year} {2005})\BibitemShut {NoStop}%
\bibitem [{\citenamefont {Aaronson}\ and\ \citenamefont
  {Gottesman}(2004)}]{Aaronson2004Improved}%
  \BibitemOpen
  \bibfield  {author} {\bibinfo {author} {\bibfnamefont {S.}~\bibnamefont
  {Aaronson}}\ and\ \bibinfo {author} {\bibfnamefont {D.}~\bibnamefont
  {Gottesman}},\ }\bibfield  {title} {\bibinfo {title} {Improved simulation of
  stabilizer circuits},\ }\href {https://doi.org/10.1103/PhysRevA.70.052328}
  {\bibfield  {journal} {\bibinfo  {journal} {Phys. Rev. A}\ }\textbf {\bibinfo
  {volume} {70}},\ \bibinfo {pages} {052328} (\bibinfo {year}
  {2004})}\BibitemShut {NoStop}%
\bibitem [{\citenamefont {Gottesman}(1998)}]{gottesman1998heisenberg}%
  \BibitemOpen
  \bibfield  {author} {\bibinfo {author} {\bibfnamefont {D.}~\bibnamefont
  {Gottesman}},\ }\href {https://arxiv.org/abs/quant-ph/9807006} {\bibinfo
  {title} {The heisenberg representation of quantum computers}} (\bibinfo
  {year} {1998}),\ \Eprint {https://arxiv.org/abs/quant-ph/9807006}
  {arXiv:quant-ph/9807006 [quant-ph]} \BibitemShut {NoStop}%
\bibitem [{\citenamefont {Nielsen}\ and\ \citenamefont
  {Chuang}(2010)}]{nielsen2010quantum}%
  \BibitemOpen
  \bibfield  {author} {\bibinfo {author} {\bibfnamefont {M.~A.}\ \bibnamefont
  {Nielsen}}\ and\ \bibinfo {author} {\bibfnamefont {I.~L.}\ \bibnamefont
  {Chuang}},\ }\href@noop {} {\emph {\bibinfo {title} {Quantum computation and
  quantum information}}}\ (\bibinfo  {publisher} {Cambridge university press},\
  \bibinfo {year} {2010})\BibitemShut {NoStop}%
\bibitem [{\citenamefont {Liu}\ and\ \citenamefont
  {Winter}(2022)}]{Liu2022Many}%
  \BibitemOpen
  \bibfield  {author} {\bibinfo {author} {\bibfnamefont {Z.-W.}\ \bibnamefont
  {Liu}}\ and\ \bibinfo {author} {\bibfnamefont {A.}~\bibnamefont {Winter}},\
  }\bibfield  {title} {\bibinfo {title} {Many-body quantum magic},\ }\href
  {https://doi.org/10.1103/PRXQuantum.3.020333} {\bibfield  {journal} {\bibinfo
   {journal} {PRX Quantum}\ }\textbf {\bibinfo {volume} {3}},\ \bibinfo {pages}
  {020333} (\bibinfo {year} {2022})}\BibitemShut {NoStop}%
\bibitem [{\citenamefont {Heinrich}\ and\ \citenamefont
  {Gross}(2019)}]{Heinrich2019robustnessofmagic}%
  \BibitemOpen
  \bibfield  {author} {\bibinfo {author} {\bibfnamefont {M.}~\bibnamefont
  {Heinrich}}\ and\ \bibinfo {author} {\bibfnamefont {D.}~\bibnamefont
  {Gross}},\ }\bibfield  {title} {\bibinfo {title} {Robustness of {M}agic and
  {S}ymmetries of the {S}tabiliser {P}olytope},\ }\href
  {https://doi.org/10.22331/q-2019-04-08-132} {\bibfield  {journal} {\bibinfo
  {journal} {{Quantum}}\ }\textbf {\bibinfo {volume} {3}},\ \bibinfo {pages}
  {132} (\bibinfo {year} {2019})}\BibitemShut {NoStop}%
\bibitem [{\citenamefont {Liu}\ \emph {et~al.}(2019)\citenamefont {Liu},
  \citenamefont {Bu},\ and\ \citenamefont {Takagi}}]{Liu2019OneShot}%
  \BibitemOpen
  \bibfield  {author} {\bibinfo {author} {\bibfnamefont {Z.-W.}\ \bibnamefont
  {Liu}}, \bibinfo {author} {\bibfnamefont {K.}~\bibnamefont {Bu}},\ and\
  \bibinfo {author} {\bibfnamefont {R.}~\bibnamefont {Takagi}},\ }\bibfield
  {title} {\bibinfo {title} {One-shot operational quantum resource theory},\
  }\href {https://doi.org/10.1103/PhysRevLett.123.020401} {\bibfield  {journal}
  {\bibinfo  {journal} {Phys. Rev. Lett.}\ }\textbf {\bibinfo {volume} {123}},\
  \bibinfo {pages} {020401} (\bibinfo {year} {2019})}\BibitemShut {NoStop}%
\bibitem [{\citenamefont {Wang}\ \emph {et~al.}(2020)\citenamefont {Wang},
  \citenamefont {Wilde},\ and\ \citenamefont {Su}}]{Wang2020Efficiently}%
  \BibitemOpen
  \bibfield  {author} {\bibinfo {author} {\bibfnamefont {X.}~\bibnamefont
  {Wang}}, \bibinfo {author} {\bibfnamefont {M.~M.}\ \bibnamefont {Wilde}},\
  and\ \bibinfo {author} {\bibfnamefont {Y.}~\bibnamefont {Su}},\ }\bibfield
  {title} {\bibinfo {title} {Efficiently computable bounds for magic state
  distillation},\ }\href {https://doi.org/10.1103/PhysRevLett.124.090505}
  {\bibfield  {journal} {\bibinfo  {journal} {Phys. Rev. Lett.}\ }\textbf
  {\bibinfo {volume} {124}},\ \bibinfo {pages} {090505} (\bibinfo {year}
  {2020})}\BibitemShut {NoStop}%
\bibitem [{\citenamefont {Heimendahl}\ \emph {et~al.}(2021)\citenamefont
  {Heimendahl}, \citenamefont {Montealegre-Mora}, \citenamefont {Vallentin},\
  and\ \citenamefont {Gross}}]{Heimendahl2021stabilizerextentis}%
  \BibitemOpen
  \bibfield  {author} {\bibinfo {author} {\bibfnamefont {A.}~\bibnamefont
  {Heimendahl}}, \bibinfo {author} {\bibfnamefont {F.}~\bibnamefont
  {Montealegre-Mora}}, \bibinfo {author} {\bibfnamefont {F.}~\bibnamefont
  {Vallentin}},\ and\ \bibinfo {author} {\bibfnamefont {D.}~\bibnamefont
  {Gross}},\ }\bibfield  {title} {\bibinfo {title} {Stabilizer extent is not
  multiplicative},\ }\href {https://doi.org/10.22331/q-2021-02-24-400}
  {\bibfield  {journal} {\bibinfo  {journal} {{Quantum}}\ }\textbf {\bibinfo
  {volume} {5}},\ \bibinfo {pages} {400} (\bibinfo {year} {2021})}\BibitemShut
  {NoStop}%
\bibitem [{\citenamefont {Veitch}\ \emph {et~al.}(2014)\citenamefont {Veitch},
  \citenamefont {Mousavian}, \citenamefont {Gottesman},\ and\ \citenamefont
  {Emerson}}]{Veitch2014resource}%
  \BibitemOpen
  \bibfield  {author} {\bibinfo {author} {\bibfnamefont {V.}~\bibnamefont
  {Veitch}}, \bibinfo {author} {\bibfnamefont {S.~A.~H.}\ \bibnamefont
  {Mousavian}}, \bibinfo {author} {\bibfnamefont {D.}~\bibnamefont
  {Gottesman}},\ and\ \bibinfo {author} {\bibfnamefont {J.}~\bibnamefont
  {Emerson}},\ }\bibfield  {title} {\bibinfo {title} {The resource theory of
  stabilizer quantum computation},\ }\href
  {https://doi.org/10.1088/1367-2630/16/1/013009} {\bibfield  {journal}
  {\bibinfo  {journal} {New Journal of Physics}\ }\textbf {\bibinfo {volume}
  {16}},\ \bibinfo {pages} {013009} (\bibinfo {year} {2014})}\BibitemShut
  {NoStop}%
\bibitem [{\citenamefont {Seddon}\ and\ \citenamefont
  {Campbell}(2019)}]{seddon2019quantifying}%
  \BibitemOpen
  \bibfield  {author} {\bibinfo {author} {\bibfnamefont {J.~R.}\ \bibnamefont
  {Seddon}}\ and\ \bibinfo {author} {\bibfnamefont {E.~T.}\ \bibnamefont
  {Campbell}},\ }\bibfield  {title} {\bibinfo {title} {Quantifying magic for
  multi-qubit operations},\ }\href {https://doi.org/10.1098/rspa.2019.0251}
  {\bibfield  {journal} {\bibinfo  {journal} {Proceedings of the Royal Society
  A}\ }\textbf {\bibinfo {volume} {475}},\ \bibinfo {pages} {20190251}
  (\bibinfo {year} {2019})}\BibitemShut {NoStop}%
\bibitem [{\citenamefont {Ghosh}\ \emph {et~al.}(2023)\citenamefont {Ghosh},
  \citenamefont {Deshpande}, \citenamefont {Hangleiter}, \citenamefont
  {Gorshkov},\ and\ \citenamefont {Fefferman}}]{Ghosh2015Complexity}%
  \BibitemOpen
  \bibfield  {author} {\bibinfo {author} {\bibfnamefont {S.}~\bibnamefont
  {Ghosh}}, \bibinfo {author} {\bibfnamefont {A.}~\bibnamefont {Deshpande}},
  \bibinfo {author} {\bibfnamefont {D.}~\bibnamefont {Hangleiter}}, \bibinfo
  {author} {\bibfnamefont {A.~V.}\ \bibnamefont {Gorshkov}},\ and\ \bibinfo
  {author} {\bibfnamefont {B.}~\bibnamefont {Fefferman}},\ }\bibfield  {title}
  {\bibinfo {title} {Complexity phase transitions generated by entanglement},\
  }\href {https://doi.org/10.1103/PhysRevLett.131.030601} {\bibfield  {journal}
  {\bibinfo  {journal} {Phys. Rev. Lett.}\ }\textbf {\bibinfo {volume} {131}},\
  \bibinfo {pages} {030601} (\bibinfo {year} {2023})}\BibitemShut {NoStop}%
\bibitem [{\citenamefont {Skinner}\ \emph {et~al.}(2019)\citenamefont
  {Skinner}, \citenamefont {Ruhman},\ and\ \citenamefont
  {Nahum}}]{Skinner_2019}%
  \BibitemOpen
  \bibfield  {author} {\bibinfo {author} {\bibfnamefont {B.}~\bibnamefont
  {Skinner}}, \bibinfo {author} {\bibfnamefont {J.}~\bibnamefont {Ruhman}},\
  and\ \bibinfo {author} {\bibfnamefont {A.}~\bibnamefont {Nahum}},\ }\bibfield
   {title} {\bibinfo {title} {Measurement-induced phase transitions in the
  dynamics of entanglement},\ }\href
  {https://doi.org/10.1103/PhysRevX.9.031009} {\bibfield  {journal} {\bibinfo
  {journal} {Phys. Rev. X}\ }\textbf {\bibinfo {volume} {9}},\ \bibinfo {pages}
  {031009} (\bibinfo {year} {2019})}\BibitemShut {NoStop}%
\bibitem [{\citenamefont {Koh}\ \emph {et~al.}(2023)\citenamefont {Koh},
  \citenamefont {Sun}, \citenamefont {Motta},\ and\ \citenamefont
  {Minnich}}]{koh2023measurement}%
  \BibitemOpen
  \bibfield  {author} {\bibinfo {author} {\bibfnamefont {J.~M.}\ \bibnamefont
  {Koh}}, \bibinfo {author} {\bibfnamefont {S.-N.}\ \bibnamefont {Sun}},
  \bibinfo {author} {\bibfnamefont {M.}~\bibnamefont {Motta}},\ and\ \bibinfo
  {author} {\bibfnamefont {A.~J.}\ \bibnamefont {Minnich}},\ }\bibfield
  {title} {\bibinfo {title} {Measurement-induced entanglement phase transition
  on a superconducting quantum processor with mid-circuit readout},\ }\href
  {https://doi.org/10.1038/s41567-023-02076-6} {\bibfield  {journal} {\bibinfo
  {journal} {Nature Physics}\ }\textbf {\bibinfo {volume} {19}},\ \bibinfo
  {pages} {1314} (\bibinfo {year} {2023})}\BibitemShut {NoStop}%
\bibitem [{\citenamefont {Choi}\ \emph {et~al.}(2020)\citenamefont {Choi},
  \citenamefont {Bao}, \citenamefont {Qi},\ and\ \citenamefont
  {Altman}}]{Choi_2020}%
  \BibitemOpen
  \bibfield  {author} {\bibinfo {author} {\bibfnamefont {S.}~\bibnamefont
  {Choi}}, \bibinfo {author} {\bibfnamefont {Y.}~\bibnamefont {Bao}}, \bibinfo
  {author} {\bibfnamefont {X.-L.}\ \bibnamefont {Qi}},\ and\ \bibinfo {author}
  {\bibfnamefont {E.}~\bibnamefont {Altman}},\ }\bibfield  {title} {\bibinfo
  {title} {Quantum error correction in scrambling dynamics and
  measurement-induced phase transition},\ }\href
  {https://doi.org/10.1103/PhysRevLett.125.030505} {\bibfield  {journal}
  {\bibinfo  {journal} {Phys. Rev. Lett.}\ }\textbf {\bibinfo {volume} {125}},\
  \bibinfo {pages} {030505} (\bibinfo {year} {2020})}\BibitemShut {NoStop}%
\bibitem [{\citenamefont {Gullans}\ and\ \citenamefont
  {Huse}(2020)}]{Gullans_2020}%
  \BibitemOpen
  \bibfield  {author} {\bibinfo {author} {\bibfnamefont {M.~J.}\ \bibnamefont
  {Gullans}}\ and\ \bibinfo {author} {\bibfnamefont {D.~A.}\ \bibnamefont
  {Huse}},\ }\bibfield  {title} {\bibinfo {title} {Dynamical purification phase
  transition induced by quantum measurements},\ }\href
  {https://doi.org/10.1103/PhysRevX.10.041020} {\bibfield  {journal} {\bibinfo
  {journal} {Phys. Rev. X}\ }\textbf {\bibinfo {volume} {10}},\ \bibinfo
  {pages} {041020} (\bibinfo {year} {2020})}\BibitemShut {NoStop}%
\bibitem [{\citenamefont {Fux}\ \emph {et~al.}(2024)\citenamefont {Fux},
  \citenamefont {Tirrito}, \citenamefont {Dalmonte},\ and\ \citenamefont
  {Fazio}}]{Fux2024Entanglement}%
  \BibitemOpen
  \bibfield  {author} {\bibinfo {author} {\bibfnamefont {G.~E.}\ \bibnamefont
  {Fux}}, \bibinfo {author} {\bibfnamefont {E.}~\bibnamefont {Tirrito}},
  \bibinfo {author} {\bibfnamefont {M.}~\bibnamefont {Dalmonte}},\ and\
  \bibinfo {author} {\bibfnamefont {R.}~\bibnamefont {Fazio}},\ }\bibfield
  {title} {\bibinfo {title} {Entanglement -- nonstabilizerness separation in
  hybrid quantum circuits},\ }\href
  {https://doi.org/10.1103/PhysRevResearch.6.L042030} {\bibfield  {journal}
  {\bibinfo  {journal} {Phys. Rev. Res.}\ }\textbf {\bibinfo {volume} {6}},\
  \bibinfo {pages} {L042030} (\bibinfo {year} {2024})}\BibitemShut {NoStop}%
\bibitem [{\citenamefont {Suzuki}\ \emph {et~al.}(2025)\citenamefont {Suzuki},
  \citenamefont {Haferkamp}, \citenamefont {Eisert},\ and\ \citenamefont
  {Faist}}]{suzuki2025quantum}%
  \BibitemOpen
  \bibfield  {author} {\bibinfo {author} {\bibfnamefont {R.}~\bibnamefont
  {Suzuki}}, \bibinfo {author} {\bibfnamefont {J.}~\bibnamefont {Haferkamp}},
  \bibinfo {author} {\bibfnamefont {J.}~\bibnamefont {Eisert}},\ and\ \bibinfo
  {author} {\bibfnamefont {P.}~\bibnamefont {Faist}},\ }\bibfield  {title}
  {\bibinfo {title} {Quantum complexity phase transitions in monitored random
  circuits},\ }\href {https://doi.org/10.22331/q-2025-02-10-1627} {\bibfield
  {journal} {\bibinfo  {journal} {Quantum}\ }\textbf {\bibinfo {volume} {9}},\
  \bibinfo {pages} {1627} (\bibinfo {year} {2025})}\BibitemShut {NoStop}%
\bibitem [{\citenamefont {Niroula}\ \emph {et~al.}(2024)\citenamefont
  {Niroula}, \citenamefont {White}, \citenamefont {Wang}, \citenamefont
  {Johri}, \citenamefont {Zhu}, \citenamefont {Monroe}, \citenamefont {Noel},\
  and\ \citenamefont {Gullans}}]{niroula2024phase}%
  \BibitemOpen
  \bibfield  {author} {\bibinfo {author} {\bibfnamefont {P.}~\bibnamefont
  {Niroula}}, \bibinfo {author} {\bibfnamefont {C.~D.}\ \bibnamefont {White}},
  \bibinfo {author} {\bibfnamefont {Q.}~\bibnamefont {Wang}}, \bibinfo {author}
  {\bibfnamefont {S.}~\bibnamefont {Johri}}, \bibinfo {author} {\bibfnamefont
  {D.}~\bibnamefont {Zhu}}, \bibinfo {author} {\bibfnamefont {C.}~\bibnamefont
  {Monroe}}, \bibinfo {author} {\bibfnamefont {C.}~\bibnamefont {Noel}},\ and\
  \bibinfo {author} {\bibfnamefont {M.~J.}\ \bibnamefont {Gullans}},\
  }\bibfield  {title} {\bibinfo {title} {Phase transition in magic with random
  quantum circuits},\ }\href {https://doi.org/10.1038/s41567-024-02637-3}
  {\bibfield  {journal} {\bibinfo  {journal} {Nature physics}\ }\textbf
  {\bibinfo {volume} {20}},\ \bibinfo {pages} {1786} (\bibinfo {year}
  {2024})}\BibitemShut {NoStop}%
\bibitem [{\citenamefont {Hashimoto}\ \emph {et~al.}(2017)\citenamefont
  {Hashimoto}, \citenamefont {Murata},\ and\ \citenamefont
  {Yoshii}}]{hashimoto2017out}%
  \BibitemOpen
  \bibfield  {author} {\bibinfo {author} {\bibfnamefont {K.}~\bibnamefont
  {Hashimoto}}, \bibinfo {author} {\bibfnamefont {K.}~\bibnamefont {Murata}},\
  and\ \bibinfo {author} {\bibfnamefont {R.}~\bibnamefont {Yoshii}},\
  }\bibfield  {title} {\bibinfo {title} {Out-of-time-order correlators in
  quantum mechanics},\ }\href {https://doi.org/10.1007/JHEP10(2017)138}
  {\bibfield  {journal} {\bibinfo  {journal} {Journal of High Energy Physics}\
  }\textbf {\bibinfo {volume} {2017}},\ \bibinfo {pages} {1} (\bibinfo {year}
  {2017})}\BibitemShut {NoStop}%
\bibitem [{\citenamefont {Swingle}(2018)}]{swingle2018unscrambling}%
  \BibitemOpen
  \bibfield  {author} {\bibinfo {author} {\bibfnamefont {B.}~\bibnamefont
  {Swingle}},\ }\bibfield  {title} {\bibinfo {title} {Unscrambling the physics
  of out-of-time-order correlators},\ }\href
  {https://doi.org/10.1038/s41567-018-0295-5} {\bibfield  {journal} {\bibinfo
  {journal} {Nature Physics}\ }\textbf {\bibinfo {volume} {14}},\ \bibinfo
  {pages} {988} (\bibinfo {year} {2018})}\BibitemShut {NoStop}%
\bibitem [{\citenamefont {Li}\ \emph {et~al.}(2017)\citenamefont {Li},
  \citenamefont {Fan}, \citenamefont {Wang}, \citenamefont {Ye}, \citenamefont
  {Zeng}, \citenamefont {Zhai}, \citenamefont {Peng},\ and\ \citenamefont
  {Du}}]{Li2017Measuring}%
  \BibitemOpen
  \bibfield  {author} {\bibinfo {author} {\bibfnamefont {J.}~\bibnamefont
  {Li}}, \bibinfo {author} {\bibfnamefont {R.}~\bibnamefont {Fan}}, \bibinfo
  {author} {\bibfnamefont {H.}~\bibnamefont {Wang}}, \bibinfo {author}
  {\bibfnamefont {B.}~\bibnamefont {Ye}}, \bibinfo {author} {\bibfnamefont
  {B.}~\bibnamefont {Zeng}}, \bibinfo {author} {\bibfnamefont {H.}~\bibnamefont
  {Zhai}}, \bibinfo {author} {\bibfnamefont {X.}~\bibnamefont {Peng}},\ and\
  \bibinfo {author} {\bibfnamefont {J.}~\bibnamefont {Du}},\ }\bibfield
  {title} {\bibinfo {title} {Measuring out-of-time-order correlators on a
  nuclear magnetic resonance quantum simulator},\ }\href
  {https://doi.org/10.1103/PhysRevX.7.031011} {\bibfield  {journal} {\bibinfo
  {journal} {Phys. Rev. X}\ }\textbf {\bibinfo {volume} {7}},\ \bibinfo {pages}
  {031011} (\bibinfo {year} {2017})}\BibitemShut {NoStop}%
\bibitem [{\citenamefont {Green}\ \emph {et~al.}(2022)\citenamefont {Green},
  \citenamefont {Elben}, \citenamefont {Alderete}, \citenamefont {Joshi},
  \citenamefont {Nguyen}, \citenamefont {Zache}, \citenamefont {Zhu},
  \citenamefont {Sundar},\ and\ \citenamefont {Linke}}]{Green2022Experimental}%
  \BibitemOpen
  \bibfield  {author} {\bibinfo {author} {\bibfnamefont {A.~M.}\ \bibnamefont
  {Green}}, \bibinfo {author} {\bibfnamefont {A.}~\bibnamefont {Elben}},
  \bibinfo {author} {\bibfnamefont {C.~H.}\ \bibnamefont {Alderete}}, \bibinfo
  {author} {\bibfnamefont {L.~K.}\ \bibnamefont {Joshi}}, \bibinfo {author}
  {\bibfnamefont {N.~H.}\ \bibnamefont {Nguyen}}, \bibinfo {author}
  {\bibfnamefont {T.~V.}\ \bibnamefont {Zache}}, \bibinfo {author}
  {\bibfnamefont {Y.}~\bibnamefont {Zhu}}, \bibinfo {author} {\bibfnamefont
  {B.}~\bibnamefont {Sundar}},\ and\ \bibinfo {author} {\bibfnamefont {N.~M.}\
  \bibnamefont {Linke}},\ }\bibfield  {title} {\bibinfo {title} {Experimental
  measurement of out-of-time-ordered correlators at finite temperature},\
  }\href {https://doi.org/10.1103/PhysRevLett.128.140601} {\bibfield  {journal}
  {\bibinfo  {journal} {Phys. Rev. Lett.}\ }\textbf {\bibinfo {volume} {128}},\
  \bibinfo {pages} {140601} (\bibinfo {year} {2022})}\BibitemShut {NoStop}%
\bibitem [{\citenamefont {Lewis-Swan}\ \emph {et~al.}(2019)\citenamefont
  {Lewis-Swan}, \citenamefont {Safavi-Naini}, \citenamefont {Kaufman},\ and\
  \citenamefont {Rey}}]{lewis2019dynamics}%
  \BibitemOpen
  \bibfield  {author} {\bibinfo {author} {\bibfnamefont {R.}~\bibnamefont
  {Lewis-Swan}}, \bibinfo {author} {\bibfnamefont {A.}~\bibnamefont
  {Safavi-Naini}}, \bibinfo {author} {\bibfnamefont {A.}~\bibnamefont
  {Kaufman}},\ and\ \bibinfo {author} {\bibfnamefont {A.}~\bibnamefont {Rey}},\
  }\bibfield  {title} {\bibinfo {title} {Dynamics of quantum information},\
  }\href {https://doi.org/10.1038/s42254-019-0090-y} {\bibfield  {journal}
  {\bibinfo  {journal} {Nature Reviews Physics}\ }\textbf {\bibinfo {volume}
  {1}},\ \bibinfo {pages} {627} (\bibinfo {year} {2019})}\BibitemShut {NoStop}%
\bibitem [{\citenamefont {Landsman}\ \emph {et~al.}(2019)\citenamefont
  {Landsman}, \citenamefont {Figgatt}, \citenamefont {Schuster}, \citenamefont
  {Linke}, \citenamefont {Yoshida}, \citenamefont {Yao},\ and\ \citenamefont
  {Monroe}}]{landsman2019verified}%
  \BibitemOpen
  \bibfield  {author} {\bibinfo {author} {\bibfnamefont {K.~A.}\ \bibnamefont
  {Landsman}}, \bibinfo {author} {\bibfnamefont {C.}~\bibnamefont {Figgatt}},
  \bibinfo {author} {\bibfnamefont {T.}~\bibnamefont {Schuster}}, \bibinfo
  {author} {\bibfnamefont {N.~M.}\ \bibnamefont {Linke}}, \bibinfo {author}
  {\bibfnamefont {B.}~\bibnamefont {Yoshida}}, \bibinfo {author} {\bibfnamefont
  {N.~Y.}\ \bibnamefont {Yao}},\ and\ \bibinfo {author} {\bibfnamefont
  {C.}~\bibnamefont {Monroe}},\ }\bibfield  {title} {\bibinfo {title} {Verified
  quantum information scrambling},\ }\href
  {https://doi.org/10.1038/s41586-019-0952-6} {\bibfield  {journal} {\bibinfo
  {journal} {Nature}\ }\textbf {\bibinfo {volume} {567}},\ \bibinfo {pages}
  {61} (\bibinfo {year} {2019})}\BibitemShut {NoStop}%
\bibitem [{\citenamefont {Mi}\ \emph {et~al.}(2021)\citenamefont {Mi},
  \citenamefont {Roushan}, \citenamefont {Quintana}, \citenamefont {Mandrà},
  \citenamefont {Marshall}, \citenamefont {Neill}, \citenamefont {Arute},
  \citenamefont {Arya}, \citenamefont {Atalaya}, \citenamefont {Babbush},
  \citenamefont {Bardin}, \citenamefont {Barends}, \citenamefont {Basso},
  \citenamefont {Bengtsson}, \citenamefont {Boixo}, \citenamefont {Bourassa},
  \citenamefont {Broughton}, \citenamefont {Buckley}, \citenamefont {Buell},
  \citenamefont {Burkett}, \citenamefont {Bushnell}, \citenamefont {Chen},
  \citenamefont {Chiaro}, \citenamefont {Collins}, \citenamefont {Courtney},
  \citenamefont {Demura}, \citenamefont {Derk}, \citenamefont {Dunsworth},
  \citenamefont {Eppens}, \citenamefont {Erickson}, \citenamefont {Farhi},
  \citenamefont {Fowler}, \citenamefont {Foxen}, \citenamefont {Gidney},
  \citenamefont {Giustina}, \citenamefont {Gross}, \citenamefont {Harrigan},
  \citenamefont {Harrington}, \citenamefont {Hilton}, \citenamefont {Ho},
  \citenamefont {Hong}, \citenamefont {Huang}, \citenamefont {Huggins},
  \citenamefont {Ioffe}, \citenamefont {Isakov}, \citenamefont {Jeffrey},
  \citenamefont {Jiang}, \citenamefont {Jones}, \citenamefont {Kafri},
  \citenamefont {Kelly}, \citenamefont {Kim}, \citenamefont {Kitaev},
  \citenamefont {Klimov}, \citenamefont {Korotkov}, \citenamefont {Kostritsa},
  \citenamefont {Landhuis}, \citenamefont {Laptev}, \citenamefont {Lucero},
  \citenamefont {Martin}, \citenamefont {McClean}, \citenamefont {McCourt},
  \citenamefont {McEwen}, \citenamefont {Megrant}, \citenamefont {Miao},
  \citenamefont {Mohseni}, \citenamefont {Montazeri}, \citenamefont
  {Mruczkiewicz}, \citenamefont {Mutus}, \citenamefont {Naaman}, \citenamefont
  {Neeley}, \citenamefont {Newman}, \citenamefont {Niu}, \citenamefont
  {O’Brien}, \citenamefont {Opremcak}, \citenamefont {Ostby}, \citenamefont
  {Pato}, \citenamefont {Petukhov}, \citenamefont {Redd}, \citenamefont
  {Rubin}, \citenamefont {Sank}, \citenamefont {Satzinger}, \citenamefont
  {Shvarts}, \citenamefont {Strain}, \citenamefont {Szalay}, \citenamefont
  {Trevithick}, \citenamefont {Villalonga}, \citenamefont {White},
  \citenamefont {Yao}, \citenamefont {Yeh}, \citenamefont {Zalcman},
  \citenamefont {Neven}, \citenamefont {Aleiner}, \citenamefont {Kechedzhi},
  \citenamefont {Smelyanskiy},\ and\ \citenamefont
  {Chen}}]{Xiao2021Information}%
  \BibitemOpen
  \bibfield  {author} {\bibinfo {author} {\bibfnamefont {X.}~\bibnamefont
  {Mi}}, \bibinfo {author} {\bibfnamefont {P.}~\bibnamefont {Roushan}},
  \bibinfo {author} {\bibfnamefont {C.}~\bibnamefont {Quintana}}, \bibinfo
  {author} {\bibfnamefont {S.}~\bibnamefont {Mandrà}}, \bibinfo {author}
  {\bibfnamefont {J.}~\bibnamefont {Marshall}}, \bibinfo {author}
  {\bibfnamefont {C.}~\bibnamefont {Neill}}, \bibinfo {author} {\bibfnamefont
  {F.}~\bibnamefont {Arute}}, \bibinfo {author} {\bibfnamefont
  {K.}~\bibnamefont {Arya}}, \bibinfo {author} {\bibfnamefont {J.}~\bibnamefont
  {Atalaya}}, \bibinfo {author} {\bibfnamefont {R.}~\bibnamefont {Babbush}},
  \bibinfo {author} {\bibfnamefont {J.~C.}\ \bibnamefont {Bardin}}, \bibinfo
  {author} {\bibfnamefont {R.}~\bibnamefont {Barends}}, \bibinfo {author}
  {\bibfnamefont {J.}~\bibnamefont {Basso}}, \bibinfo {author} {\bibfnamefont
  {A.}~\bibnamefont {Bengtsson}}, \bibinfo {author} {\bibfnamefont
  {S.}~\bibnamefont {Boixo}}, \bibinfo {author} {\bibfnamefont
  {A.}~\bibnamefont {Bourassa}}, \bibinfo {author} {\bibfnamefont
  {M.}~\bibnamefont {Broughton}}, \bibinfo {author} {\bibfnamefont {B.~B.}\
  \bibnamefont {Buckley}}, \bibinfo {author} {\bibfnamefont {D.~A.}\
  \bibnamefont {Buell}}, \bibinfo {author} {\bibfnamefont {B.}~\bibnamefont
  {Burkett}}, \bibinfo {author} {\bibfnamefont {N.}~\bibnamefont {Bushnell}},
  \bibinfo {author} {\bibfnamefont {Z.}~\bibnamefont {Chen}}, \bibinfo {author}
  {\bibfnamefont {B.}~\bibnamefont {Chiaro}}, \bibinfo {author} {\bibfnamefont
  {R.}~\bibnamefont {Collins}}, \bibinfo {author} {\bibfnamefont
  {W.}~\bibnamefont {Courtney}}, \bibinfo {author} {\bibfnamefont
  {S.}~\bibnamefont {Demura}}, \bibinfo {author} {\bibfnamefont {A.~R.}\
  \bibnamefont {Derk}}, \bibinfo {author} {\bibfnamefont {A.}~\bibnamefont
  {Dunsworth}}, \bibinfo {author} {\bibfnamefont {D.}~\bibnamefont {Eppens}},
  \bibinfo {author} {\bibfnamefont {C.}~\bibnamefont {Erickson}}, \bibinfo
  {author} {\bibfnamefont {E.}~\bibnamefont {Farhi}}, \bibinfo {author}
  {\bibfnamefont {A.~G.}\ \bibnamefont {Fowler}}, \bibinfo {author}
  {\bibfnamefont {B.}~\bibnamefont {Foxen}}, \bibinfo {author} {\bibfnamefont
  {C.}~\bibnamefont {Gidney}}, \bibinfo {author} {\bibfnamefont
  {M.}~\bibnamefont {Giustina}}, \bibinfo {author} {\bibfnamefont {J.~A.}\
  \bibnamefont {Gross}}, \bibinfo {author} {\bibfnamefont {M.~P.}\ \bibnamefont
  {Harrigan}}, \bibinfo {author} {\bibfnamefont {S.~D.}\ \bibnamefont
  {Harrington}}, \bibinfo {author} {\bibfnamefont {J.}~\bibnamefont {Hilton}},
  \bibinfo {author} {\bibfnamefont {A.}~\bibnamefont {Ho}}, \bibinfo {author}
  {\bibfnamefont {S.}~\bibnamefont {Hong}}, \bibinfo {author} {\bibfnamefont
  {T.}~\bibnamefont {Huang}}, \bibinfo {author} {\bibfnamefont {W.~J.}\
  \bibnamefont {Huggins}}, \bibinfo {author} {\bibfnamefont {L.~B.}\
  \bibnamefont {Ioffe}}, \bibinfo {author} {\bibfnamefont {S.~V.}\ \bibnamefont
  {Isakov}}, \bibinfo {author} {\bibfnamefont {E.}~\bibnamefont {Jeffrey}},
  \bibinfo {author} {\bibfnamefont {Z.}~\bibnamefont {Jiang}}, \bibinfo
  {author} {\bibfnamefont {C.}~\bibnamefont {Jones}}, \bibinfo {author}
  {\bibfnamefont {D.}~\bibnamefont {Kafri}}, \bibinfo {author} {\bibfnamefont
  {J.}~\bibnamefont {Kelly}}, \bibinfo {author} {\bibfnamefont
  {S.}~\bibnamefont {Kim}}, \bibinfo {author} {\bibfnamefont {A.}~\bibnamefont
  {Kitaev}}, \bibinfo {author} {\bibfnamefont {P.~V.}\ \bibnamefont {Klimov}},
  \bibinfo {author} {\bibfnamefont {A.~N.}\ \bibnamefont {Korotkov}}, \bibinfo
  {author} {\bibfnamefont {F.}~\bibnamefont {Kostritsa}}, \bibinfo {author}
  {\bibfnamefont {D.}~\bibnamefont {Landhuis}}, \bibinfo {author}
  {\bibfnamefont {P.}~\bibnamefont {Laptev}}, \bibinfo {author} {\bibfnamefont
  {E.}~\bibnamefont {Lucero}}, \bibinfo {author} {\bibfnamefont
  {O.}~\bibnamefont {Martin}}, \bibinfo {author} {\bibfnamefont {J.~R.}\
  \bibnamefont {McClean}}, \bibinfo {author} {\bibfnamefont {T.}~\bibnamefont
  {McCourt}}, \bibinfo {author} {\bibfnamefont {M.}~\bibnamefont {McEwen}},
  \bibinfo {author} {\bibfnamefont {A.}~\bibnamefont {Megrant}}, \bibinfo
  {author} {\bibfnamefont {K.~C.}\ \bibnamefont {Miao}}, \bibinfo {author}
  {\bibfnamefont {M.}~\bibnamefont {Mohseni}}, \bibinfo {author} {\bibfnamefont
  {S.}~\bibnamefont {Montazeri}}, \bibinfo {author} {\bibfnamefont
  {W.}~\bibnamefont {Mruczkiewicz}}, \bibinfo {author} {\bibfnamefont
  {J.}~\bibnamefont {Mutus}}, \bibinfo {author} {\bibfnamefont
  {O.}~\bibnamefont {Naaman}}, \bibinfo {author} {\bibfnamefont
  {M.}~\bibnamefont {Neeley}}, \bibinfo {author} {\bibfnamefont
  {M.}~\bibnamefont {Newman}}, \bibinfo {author} {\bibfnamefont {M.~Y.}\
  \bibnamefont {Niu}}, \bibinfo {author} {\bibfnamefont {T.~E.}\ \bibnamefont
  {O’Brien}}, \bibinfo {author} {\bibfnamefont {A.}~\bibnamefont {Opremcak}},
  \bibinfo {author} {\bibfnamefont {E.}~\bibnamefont {Ostby}}, \bibinfo
  {author} {\bibfnamefont {B.}~\bibnamefont {Pato}}, \bibinfo {author}
  {\bibfnamefont {A.}~\bibnamefont {Petukhov}}, \bibinfo {author}
  {\bibfnamefont {N.}~\bibnamefont {Redd}}, \bibinfo {author} {\bibfnamefont
  {N.~C.}\ \bibnamefont {Rubin}}, \bibinfo {author} {\bibfnamefont
  {D.}~\bibnamefont {Sank}}, \bibinfo {author} {\bibfnamefont {K.~J.}\
  \bibnamefont {Satzinger}}, \bibinfo {author} {\bibfnamefont {V.}~\bibnamefont
  {Shvarts}}, \bibinfo {author} {\bibfnamefont {D.}~\bibnamefont {Strain}},
  \bibinfo {author} {\bibfnamefont {M.}~\bibnamefont {Szalay}}, \bibinfo
  {author} {\bibfnamefont {M.~D.}\ \bibnamefont {Trevithick}}, \bibinfo
  {author} {\bibfnamefont {B.}~\bibnamefont {Villalonga}}, \bibinfo {author}
  {\bibfnamefont {T.}~\bibnamefont {White}}, \bibinfo {author} {\bibfnamefont
  {Z.~J.}\ \bibnamefont {Yao}}, \bibinfo {author} {\bibfnamefont
  {P.}~\bibnamefont {Yeh}}, \bibinfo {author} {\bibfnamefont {A.}~\bibnamefont
  {Zalcman}}, \bibinfo {author} {\bibfnamefont {H.}~\bibnamefont {Neven}},
  \bibinfo {author} {\bibfnamefont {I.}~\bibnamefont {Aleiner}}, \bibinfo
  {author} {\bibfnamefont {K.}~\bibnamefont {Kechedzhi}}, \bibinfo {author}
  {\bibfnamefont {V.}~\bibnamefont {Smelyanskiy}},\ and\ \bibinfo {author}
  {\bibfnamefont {Y.}~\bibnamefont {Chen}},\ }\bibfield  {title} {\bibinfo
  {title} {Information scrambling in quantum circuits},\ }\href
  {https://doi.org/10.1126/science.abg5029} {\bibfield  {journal} {\bibinfo
  {journal} {Science}\ }\textbf {\bibinfo {volume} {374}},\ \bibinfo {pages}
  {1479} (\bibinfo {year} {2021})},\ \Eprint
  {https://arxiv.org/abs/https://www.science.org/doi/pdf/10.1126/science.abg5029}
  {https://www.science.org/doi/pdf/10.1126/science.abg5029} \BibitemShut
  {NoStop}%
\bibitem [{\citenamefont {Parker}\ \emph {et~al.}(2019)\citenamefont {Parker},
  \citenamefont {Cao}, \citenamefont {Avdoshkin}, \citenamefont {Scaffidi},\
  and\ \citenamefont {Altman}}]{Parker2019Universal}%
  \BibitemOpen
  \bibfield  {author} {\bibinfo {author} {\bibfnamefont {D.~E.}\ \bibnamefont
  {Parker}}, \bibinfo {author} {\bibfnamefont {X.}~\bibnamefont {Cao}},
  \bibinfo {author} {\bibfnamefont {A.}~\bibnamefont {Avdoshkin}}, \bibinfo
  {author} {\bibfnamefont {T.}~\bibnamefont {Scaffidi}},\ and\ \bibinfo
  {author} {\bibfnamefont {E.}~\bibnamefont {Altman}},\ }\bibfield  {title}
  {\bibinfo {title} {A universal operator growth hypothesis},\ }\href
  {https://doi.org/10.1103/PhysRevX.9.041017} {\bibfield  {journal} {\bibinfo
  {journal} {Phys. Rev. X}\ }\textbf {\bibinfo {volume} {9}},\ \bibinfo {pages}
  {041017} (\bibinfo {year} {2019})}\BibitemShut {NoStop}%
\bibitem [{\citenamefont {Bhattacharjee}\ \emph {et~al.}(2022)\citenamefont
  {Bhattacharjee}, \citenamefont {Cao}, \citenamefont {Nandy},\ and\
  \citenamefont {Pathak}}]{bhattacharjee2022krylov}%
  \BibitemOpen
  \bibfield  {author} {\bibinfo {author} {\bibfnamefont {B.}~\bibnamefont
  {Bhattacharjee}}, \bibinfo {author} {\bibfnamefont {X.}~\bibnamefont {Cao}},
  \bibinfo {author} {\bibfnamefont {P.}~\bibnamefont {Nandy}},\ and\ \bibinfo
  {author} {\bibfnamefont {T.}~\bibnamefont {Pathak}},\ }\bibfield  {title}
  {\bibinfo {title} {Krylov complexity in saddle-dominated scrambling},\ }\href
  {https://doi.org/10.1007/JHEP05(2022)174} {\bibfield  {journal} {\bibinfo
  {journal} {Journal of High Energy Physics}\ }\textbf {\bibinfo {volume}
  {2022}},\ \bibinfo {pages} {1} (\bibinfo {year} {2022})}\BibitemShut
  {NoStop}%
\bibitem [{\citenamefont {Rabinovici}\ \emph {et~al.}(2021)\citenamefont
  {Rabinovici}, \citenamefont {S{\'a}nchez-Garrido}, \citenamefont {Shir},\
  and\ \citenamefont {Sonner}}]{rabinovici2021operator}%
  \BibitemOpen
  \bibfield  {author} {\bibinfo {author} {\bibfnamefont {E.}~\bibnamefont
  {Rabinovici}}, \bibinfo {author} {\bibfnamefont {A.}~\bibnamefont
  {S{\'a}nchez-Garrido}}, \bibinfo {author} {\bibfnamefont {R.}~\bibnamefont
  {Shir}},\ and\ \bibinfo {author} {\bibfnamefont {J.}~\bibnamefont {Sonner}},\
  }\bibfield  {title} {\bibinfo {title} {Operator complexity: a journey to the
  edge of krylov space},\ }\href {https://doi.org/10.1007/JHEP06(2021)062}
  {\bibfield  {journal} {\bibinfo  {journal} {Journal of High Energy Physics}\
  }\textbf {\bibinfo {volume} {2021}},\ \bibinfo {pages} {1} (\bibinfo {year}
  {2021})}\BibitemShut {NoStop}%
\bibitem [{\citenamefont {Rabinovici}\ \emph {et~al.}(2022)\citenamefont
  {Rabinovici}, \citenamefont {S{\'a}nchez-Garrido}, \citenamefont {Shir},\
  and\ \citenamefont {Sonner}}]{rabinovici2022krylov}%
  \BibitemOpen
  \bibfield  {author} {\bibinfo {author} {\bibfnamefont {E.}~\bibnamefont
  {Rabinovici}}, \bibinfo {author} {\bibfnamefont {A.}~\bibnamefont
  {S{\'a}nchez-Garrido}}, \bibinfo {author} {\bibfnamefont {R.}~\bibnamefont
  {Shir}},\ and\ \bibinfo {author} {\bibfnamefont {J.}~\bibnamefont {Sonner}},\
  }\bibfield  {title} {\bibinfo {title} {Krylov complexity from integrability
  to chaos},\ }\href {https://doi.org/10.1007/JHEP07(2022)151} {\bibfield
  {journal} {\bibinfo  {journal} {Journal of High Energy Physics}\ }\textbf
  {\bibinfo {volume} {2022}},\ \bibinfo {pages} {1} (\bibinfo {year}
  {2022})}\BibitemShut {NoStop}%
\bibitem [{\citenamefont {Hashimoto}\ \emph {et~al.}(2023)\citenamefont
  {Hashimoto}, \citenamefont {Murata}, \citenamefont {Tanahashi},\ and\
  \citenamefont {Watanabe}}]{hashimoto2023krylov}%
  \BibitemOpen
  \bibfield  {author} {\bibinfo {author} {\bibfnamefont {K.}~\bibnamefont
  {Hashimoto}}, \bibinfo {author} {\bibfnamefont {K.}~\bibnamefont {Murata}},
  \bibinfo {author} {\bibfnamefont {N.}~\bibnamefont {Tanahashi}},\ and\
  \bibinfo {author} {\bibfnamefont {R.}~\bibnamefont {Watanabe}},\ }\bibfield
  {title} {\bibinfo {title} {Krylov complexity and chaos in quantum
  mechanics},\ }\href {https://doi.org/10.1007/JHEP11(2023)040} {\bibfield
  {journal} {\bibinfo  {journal} {Journal of High Energy Physics}\ }\textbf
  {\bibinfo {volume} {2023}},\ \bibinfo {pages} {1} (\bibinfo {year}
  {2023})}\BibitemShut {NoStop}%
\bibitem [{\citenamefont {Trigueros}\ and\ \citenamefont
  {Lin}(2022)}]{Fabian2022Krylov}%
  \BibitemOpen
  \bibfield  {author} {\bibinfo {author} {\bibfnamefont {F.~B.}\ \bibnamefont
  {Trigueros}}\ and\ \bibinfo {author} {\bibfnamefont {C.-J.}\ \bibnamefont
  {Lin}},\ }\bibfield  {title} {\bibinfo {title} {{Krylov complexity of
  many-body localization: Operator localization in Krylov basis}},\ }\href
  {https://doi.org/10.21468/SciPostPhys.13.2.037} {\bibfield  {journal}
  {\bibinfo  {journal} {SciPost Phys.}\ }\textbf {\bibinfo {volume} {13}},\
  \bibinfo {pages} {037} (\bibinfo {year} {2022})}\BibitemShut {NoStop}%
\bibitem [{\citenamefont {Avdoshkin}\ \emph {et~al.}(2022)\citenamefont
  {Avdoshkin}, \citenamefont {Dymarsky},\ and\ \citenamefont
  {Smolkin}}]{avdoshkin2022krylov}%
  \BibitemOpen
  \bibfield  {author} {\bibinfo {author} {\bibfnamefont {A.}~\bibnamefont
  {Avdoshkin}}, \bibinfo {author} {\bibfnamefont {A.}~\bibnamefont
  {Dymarsky}},\ and\ \bibinfo {author} {\bibfnamefont {M.}~\bibnamefont
  {Smolkin}},\ }\href@noop {} {\bibinfo {title} {Krylov complexity in quantum
  field theory, and beyond}} (\bibinfo {year} {2022}),\ \Eprint
  {https://arxiv.org/abs/2212.14429} {arXiv:2212.14429 [hep-th]} \BibitemShut
  {NoStop}%
\bibitem [{\citenamefont {Kundu}\ \emph {et~al.}(2023)\citenamefont {Kundu},
  \citenamefont {Malvimat},\ and\ \citenamefont {Sinha}}]{kundu2023state}%
  \BibitemOpen
  \bibfield  {author} {\bibinfo {author} {\bibfnamefont {A.}~\bibnamefont
  {Kundu}}, \bibinfo {author} {\bibfnamefont {V.}~\bibnamefont {Malvimat}},\
  and\ \bibinfo {author} {\bibfnamefont {R.}~\bibnamefont {Sinha}},\ }\bibfield
   {title} {\bibinfo {title} {State dependence of krylov complexity in 2d
  cfts},\ }\href {https://doi.org/10.1007/JHEP09(2023)011} {\bibfield
  {journal} {\bibinfo  {journal} {Journal of High Energy Physics}\ }\textbf
  {\bibinfo {volume} {2023}},\ \bibinfo {pages} {1} (\bibinfo {year}
  {2023})}\BibitemShut {NoStop}%
\bibitem [{\citenamefont {Bhattacharyya}\ \emph {et~al.}(2023)\citenamefont
  {Bhattacharyya}, \citenamefont {Ghosh},\ and\ \citenamefont
  {Nandi}}]{bhattacharyya2023operator}%
  \BibitemOpen
  \bibfield  {author} {\bibinfo {author} {\bibfnamefont {A.}~\bibnamefont
  {Bhattacharyya}}, \bibinfo {author} {\bibfnamefont {D.}~\bibnamefont
  {Ghosh}},\ and\ \bibinfo {author} {\bibfnamefont {P.}~\bibnamefont {Nandi}},\
  }\bibfield  {title} {\bibinfo {title} {Operator growth and krylov complexity
  in bose-hubbard model},\ }\href {https://doi.org/10.1007/JHEP12(2023)112}
  {\bibfield  {journal} {\bibinfo  {journal} {Journal of High Energy Physics}\
  }\textbf {\bibinfo {volume} {2023}},\ \bibinfo {pages} {1} (\bibinfo {year}
  {2023})}\BibitemShut {NoStop}%
\bibitem [{\citenamefont {Caputa}\ \emph {et~al.}(2024)\citenamefont {Caputa},
  \citenamefont {Jeong}, \citenamefont {Liu}, \citenamefont {Pedraza},\ and\
  \citenamefont {Qu}}]{caputa2024krylov}%
  \BibitemOpen
  \bibfield  {author} {\bibinfo {author} {\bibfnamefont {P.}~\bibnamefont
  {Caputa}}, \bibinfo {author} {\bibfnamefont {H.-S.}\ \bibnamefont {Jeong}},
  \bibinfo {author} {\bibfnamefont {S.}~\bibnamefont {Liu}}, \bibinfo {author}
  {\bibfnamefont {J.~F.}\ \bibnamefont {Pedraza}},\ and\ \bibinfo {author}
  {\bibfnamefont {L.-C.}\ \bibnamefont {Qu}},\ }\bibfield  {title} {\bibinfo
  {title} {Krylov complexity of density matrix operators},\ }\href
  {https://doi.org/10.1007/JHEP05(2024)337} {\bibfield  {journal} {\bibinfo
  {journal} {Journal of High Energy Physics}\ }\textbf {\bibinfo {volume}
  {2024}},\ \bibinfo {pages} {1} (\bibinfo {year} {2024})}\BibitemShut
  {NoStop}%
\bibitem [{\citenamefont {Bento}\ \emph {et~al.}(2024)\citenamefont {Bento},
  \citenamefont {del Campo},\ and\ \citenamefont {C\'eleri}}]{Bento2024Krylov}%
  \BibitemOpen
  \bibfield  {author} {\bibinfo {author} {\bibfnamefont {P.~H.~S.}\
  \bibnamefont {Bento}}, \bibinfo {author} {\bibfnamefont {A.}~\bibnamefont
  {del Campo}},\ and\ \bibinfo {author} {\bibfnamefont {L.~C.}\ \bibnamefont
  {C\'eleri}},\ }\bibfield  {title} {\bibinfo {title} {Krylov complexity and
  dynamical phase transition in the quenched lipkin-meshkov-glick model},\
  }\href {https://doi.org/10.1103/PhysRevB.109.224304} {\bibfield  {journal}
  {\bibinfo  {journal} {Phys. Rev. B}\ }\textbf {\bibinfo {volume} {109}},\
  \bibinfo {pages} {224304} (\bibinfo {year} {2024})}\BibitemShut {NoStop}%
\bibitem [{\citenamefont {Stauffer}\ and\ \citenamefont
  {Aharony}(2018)}]{stauffer2018introduction}%
  \BibitemOpen
  \bibfield  {author} {\bibinfo {author} {\bibfnamefont {D.}~\bibnamefont
  {Stauffer}}\ and\ \bibinfo {author} {\bibfnamefont {A.}~\bibnamefont
  {Aharony}},\ }\href@noop {} {\emph {\bibinfo {title} {Introduction to
  percolation theory}}}\ (\bibinfo  {publisher} {Taylor \& Francis},\ \bibinfo
  {year} {2018})\BibitemShut {NoStop}%
\bibitem [{\citenamefont {Essam}(1980)}]{essam1980percolation}%
  \BibitemOpen
  \bibfield  {author} {\bibinfo {author} {\bibfnamefont {J.~W.}\ \bibnamefont
  {Essam}},\ }\bibfield  {title} {\bibinfo {title} {Percolation theory},\
  }\href {https://doi.org/10.1088/0034-4885/43/7/001} {\bibfield  {journal}
  {\bibinfo  {journal} {Reports on Progress in Physics}\ }\textbf {\bibinfo
  {volume} {43}},\ \bibinfo {pages} {833} (\bibinfo {year} {1980})}\BibitemShut
  {NoStop}%
\bibitem [{\citenamefont {Shante}\ and\ \citenamefont
  {Kirkpatrick}(1971)}]{Shante01051971}%
  \BibitemOpen
  \bibfield  {author} {\bibinfo {author} {\bibfnamefont {V.~K.}\ \bibnamefont
  {Shante}}\ and\ \bibinfo {author} {\bibfnamefont {S.}~\bibnamefont
  {Kirkpatrick}},\ }\bibfield  {title} {\bibinfo {title} {An introduction to
  percolation theory},\ }\href {https://doi.org/10.1080/00018737100101261}
  {\bibfield  {journal} {\bibinfo  {journal} {Advances in Physics}\ }\textbf
  {\bibinfo {volume} {20}},\ \bibinfo {pages} {325} (\bibinfo {year} {1971})},\
  \Eprint {https://arxiv.org/abs/https://doi.org/10.1080/00018737100101261}
  {https://doi.org/10.1080/00018737100101261} \BibitemShut {NoStop}%
\bibitem [{\citenamefont {Feng}\ \emph {et~al.}(2023)\citenamefont {Feng},
  \citenamefont {Wu}, \citenamefont {Tang}, \citenamefont {Qiao}, \citenamefont
  {Wang}, \citenamefont {Xu}, \citenamefont {Jiao}, \citenamefont {Gao},\ and\
  \citenamefont {Jin}}]{Zhen2023Direct}%
  \BibitemOpen
  \bibfield  {author} {\bibinfo {author} {\bibfnamefont {Z.}~\bibnamefont
  {Feng}}, \bibinfo {author} {\bibfnamefont {B.-H.}\ \bibnamefont {Wu}},
  \bibinfo {author} {\bibfnamefont {H.}~\bibnamefont {Tang}}, \bibinfo {author}
  {\bibfnamefont {L.-F.}\ \bibnamefont {Qiao}}, \bibinfo {author}
  {\bibfnamefont {X.-W.}\ \bibnamefont {Wang}}, \bibinfo {author}
  {\bibfnamefont {X.-Y.}\ \bibnamefont {Xu}}, \bibinfo {author} {\bibfnamefont
  {Z.-Q.}\ \bibnamefont {Jiao}}, \bibinfo {author} {\bibfnamefont
  {J.}~\bibnamefont {Gao}},\ and\ \bibinfo {author} {\bibfnamefont {X.-M.}\
  \bibnamefont {Jin}},\ }\bibfield  {title} {\bibinfo {title} {Direct
  observation of quantum percolation dynamics},\ }\href
  {https://doi.org/doi:10.1515/nanoph-2022-0324} {\bibfield  {journal}
  {\bibinfo  {journal} {Nanophotonics}\ }\textbf {\bibinfo {volume} {12}},\
  \bibinfo {pages} {559} (\bibinfo {year} {2023})}\BibitemShut {NoStop}%
\bibitem [{\citenamefont {Mookerjee}\ \emph {et~al.}(1995)\citenamefont
  {Mookerjee}, \citenamefont {Dasgupta},\ and\ \citenamefont
  {Saha}}]{mookerjee1995quantum}%
  \BibitemOpen
  \bibfield  {author} {\bibinfo {author} {\bibfnamefont {A.}~\bibnamefont
  {Mookerjee}}, \bibinfo {author} {\bibfnamefont {I.}~\bibnamefont
  {Dasgupta}},\ and\ \bibinfo {author} {\bibfnamefont {T.}~\bibnamefont
  {Saha}},\ }\bibfield  {title} {\bibinfo {title} {Quantum percolation},\
  }\href {https://doi.org/10.1142/S0217979295001129} {\bibfield  {journal}
  {\bibinfo  {journal} {International Journal of Modern Physics B}\ }\textbf
  {\bibinfo {volume} {9}},\ \bibinfo {pages} {2989} (\bibinfo {year}
  {1995})}\BibitemShut {NoStop}%
\bibitem [{\citenamefont {Shapir}\ \emph {et~al.}(1982)\citenamefont {Shapir},
  \citenamefont {Aharony},\ and\ \citenamefont
  {Harris}}]{Shapir1982Localization}%
  \BibitemOpen
  \bibfield  {author} {\bibinfo {author} {\bibfnamefont {Y.}~\bibnamefont
  {Shapir}}, \bibinfo {author} {\bibfnamefont {A.}~\bibnamefont {Aharony}},\
  and\ \bibinfo {author} {\bibfnamefont {A.~B.}\ \bibnamefont {Harris}},\
  }\bibfield  {title} {\bibinfo {title} {Localization and quantum
  percolation},\ }\href {https://doi.org/10.1103/PhysRevLett.49.486} {\bibfield
   {journal} {\bibinfo  {journal} {Phys. Rev. Lett.}\ }\textbf {\bibinfo
  {volume} {49}},\ \bibinfo {pages} {486} (\bibinfo {year} {1982})}\BibitemShut
  {NoStop}%
\bibitem [{\citenamefont {Soukoulis}\ and\ \citenamefont
  {Grest}(1991)}]{Soukoulis1991Localization}%
  \BibitemOpen
  \bibfield  {author} {\bibinfo {author} {\bibfnamefont {C.~M.}\ \bibnamefont
  {Soukoulis}}\ and\ \bibinfo {author} {\bibfnamefont {G.~S.}\ \bibnamefont
  {Grest}},\ }\bibfield  {title} {\bibinfo {title} {Localization in
  two-dimensional quantum percolation},\ }\href
  {https://doi.org/10.1103/PhysRevB.44.4685} {\bibfield  {journal} {\bibinfo
  {journal} {Phys. Rev. B}\ }\textbf {\bibinfo {volume} {44}},\ \bibinfo
  {pages} {4685} (\bibinfo {year} {1991})}\BibitemShut {NoStop}%
\bibitem [{\citenamefont {Khmel'nitskii}(1984)}]{Khmel1984Localization}%
  \BibitemOpen
  \bibfield  {author} {\bibinfo {author} {\bibfnamefont {D.}~\bibnamefont
  {Khmel'nitskii}},\ }\bibfield  {title} {\bibinfo {title} {Localization and
  coherent scattering of electrons},\ }\href
  {https://doi.org/https://doi.org/10.1016/0378-4363(84)90169-4} {\bibfield
  {journal} {\bibinfo  {journal} {Physica B+C}\ }\textbf {\bibinfo {volume}
  {126}},\ \bibinfo {pages} {235} (\bibinfo {year} {1984})}\BibitemShut
  {NoStop}%
\bibitem [{\citenamefont {Kirkpatrick}\ and\ \citenamefont
  {Eggarter}(1972)}]{Kirkpatrick1972Localized}%
  \BibitemOpen
  \bibfield  {author} {\bibinfo {author} {\bibfnamefont {S.}~\bibnamefont
  {Kirkpatrick}}\ and\ \bibinfo {author} {\bibfnamefont {T.~P.}\ \bibnamefont
  {Eggarter}},\ }\bibfield  {title} {\bibinfo {title} {Localized states of a
  binary alloy},\ }\href {https://doi.org/10.1103/PhysRevB.6.3598} {\bibfield
  {journal} {\bibinfo  {journal} {Phys. Rev. B}\ }\textbf {\bibinfo {volume}
  {6}},\ \bibinfo {pages} {3598} (\bibinfo {year} {1972})}\BibitemShut
  {NoStop}%
\bibitem [{\citenamefont {Lee}\ \emph {et~al.}(1993)\citenamefont {Lee},
  \citenamefont {Wang},\ and\ \citenamefont {Kivelson}}]{Lee1993Quantum}%
  \BibitemOpen
  \bibfield  {author} {\bibinfo {author} {\bibfnamefont {D.-H.}\ \bibnamefont
  {Lee}}, \bibinfo {author} {\bibfnamefont {Z.}~\bibnamefont {Wang}},\ and\
  \bibinfo {author} {\bibfnamefont {S.}~\bibnamefont {Kivelson}},\ }\bibfield
  {title} {\bibinfo {title} {Quantum percolation and plateau transitions in the
  quantum hall effect},\ }\href {https://doi.org/10.1103/PhysRevLett.70.4130}
  {\bibfield  {journal} {\bibinfo  {journal} {Phys. Rev. Lett.}\ }\textbf
  {\bibinfo {volume} {70}},\ \bibinfo {pages} {4130} (\bibinfo {year}
  {1993})}\BibitemShut {NoStop}%
\bibitem [{\citenamefont {Kim}\ \emph {et~al.}(2022)\citenamefont {Kim},
  \citenamefont {Murugan}, \citenamefont {Olle},\ and\ \citenamefont
  {Rosa}}]{Kim2022Operator}%
  \BibitemOpen
  \bibfield  {author} {\bibinfo {author} {\bibfnamefont {J.}~\bibnamefont
  {Kim}}, \bibinfo {author} {\bibfnamefont {J.}~\bibnamefont {Murugan}},
  \bibinfo {author} {\bibfnamefont {J.}~\bibnamefont {Olle}},\ and\ \bibinfo
  {author} {\bibfnamefont {D.}~\bibnamefont {Rosa}},\ }\bibfield  {title}
  {\bibinfo {title} {Operator delocalization in quantum networks},\ }\href
  {https://doi.org/10.1103/PhysRevA.105.L010201} {\bibfield  {journal}
  {\bibinfo  {journal} {Phys. Rev. A}\ }\textbf {\bibinfo {volume} {105}},\
  \bibinfo {pages} {L010201} (\bibinfo {year} {2022})}\BibitemShut {NoStop}%
\bibitem [{\citenamefont {Barb{\'o}n}\ \emph {et~al.}(2019)\citenamefont
  {Barb{\'o}n}, \citenamefont {Rabinovici}, \citenamefont {Shir},\ and\
  \citenamefont {Sinha}}]{barbon2019evolution}%
  \BibitemOpen
  \bibfield  {author} {\bibinfo {author} {\bibfnamefont {J.}~\bibnamefont
  {Barb{\'o}n}}, \bibinfo {author} {\bibfnamefont {E.}~\bibnamefont
  {Rabinovici}}, \bibinfo {author} {\bibfnamefont {R.}~\bibnamefont {Shir}},\
  and\ \bibinfo {author} {\bibfnamefont {R.}~\bibnamefont {Sinha}},\ }\bibfield
   {title} {\bibinfo {title} {On the evolution of operator complexity beyond
  scrambling},\ }\href {https://doi.org/10.1007/JHEP10(2019)264} {\bibfield
  {journal} {\bibinfo  {journal} {Journal of High Energy Physics}\ }\textbf
  {\bibinfo {volume} {2019}},\ \bibinfo {pages} {1} (\bibinfo {year}
  {2019})}\BibitemShut {NoStop}%
\bibitem [{\citenamefont {Dymarsky}\ and\ \citenamefont
  {Smolkin}(2021)}]{Dymarsky2021Krylov}%
  \BibitemOpen
  \bibfield  {author} {\bibinfo {author} {\bibfnamefont {A.}~\bibnamefont
  {Dymarsky}}\ and\ \bibinfo {author} {\bibfnamefont {M.}~\bibnamefont
  {Smolkin}},\ }\bibfield  {title} {\bibinfo {title} {Krylov complexity in
  conformal field theory},\ }\href
  {https://doi.org/10.1103/PhysRevD.104.L081702} {\bibfield  {journal}
  {\bibinfo  {journal} {Phys. Rev. D}\ }\textbf {\bibinfo {volume} {104}},\
  \bibinfo {pages} {L081702} (\bibinfo {year} {2021})}\BibitemShut {NoStop}%
\bibitem [{\citenamefont {Bhattacharya}\ \emph {et~al.}(2022)\citenamefont
  {Bhattacharya}, \citenamefont {Nandy}, \citenamefont {Nath},\ and\
  \citenamefont {Sahu}}]{bhattacharya2022operator}%
  \BibitemOpen
  \bibfield  {author} {\bibinfo {author} {\bibfnamefont {A.}~\bibnamefont
  {Bhattacharya}}, \bibinfo {author} {\bibfnamefont {P.}~\bibnamefont {Nandy}},
  \bibinfo {author} {\bibfnamefont {P.~P.}\ \bibnamefont {Nath}},\ and\
  \bibinfo {author} {\bibfnamefont {H.}~\bibnamefont {Sahu}},\ }\bibfield
  {title} {\bibinfo {title} {Operator growth and krylov construction in
  dissipative open quantum systems},\ }\href
  {https://doi.org/10.1007/JHEP12(2022)081} {\bibfield  {journal} {\bibinfo
  {journal} {Journal of High Energy Physics}\ }\textbf {\bibinfo {volume}
  {2022}},\ \bibinfo {pages} {1} (\bibinfo {year} {2022})}\BibitemShut
  {NoStop}%
\bibitem [{\citenamefont {Bhattacharjee}\ \emph {et~al.}(2023)\citenamefont
  {Bhattacharjee}, \citenamefont {Cao}, \citenamefont {Nandy},\ and\
  \citenamefont {Pathak}}]{bhattacharjee2023operator}%
  \BibitemOpen
  \bibfield  {author} {\bibinfo {author} {\bibfnamefont {B.}~\bibnamefont
  {Bhattacharjee}}, \bibinfo {author} {\bibfnamefont {X.}~\bibnamefont {Cao}},
  \bibinfo {author} {\bibfnamefont {P.}~\bibnamefont {Nandy}},\ and\ \bibinfo
  {author} {\bibfnamefont {T.}~\bibnamefont {Pathak}},\ }\bibfield  {title}
  {\bibinfo {title} {Operator growth in open quantum systems: lessons from the
  dissipative syk},\ }\href {https://doi.org/10.1007/JHEP03(2023)054}
  {\bibfield  {journal} {\bibinfo  {journal} {Journal of High Energy Physics}\
  }\textbf {\bibinfo {volume} {2023}},\ \bibinfo {pages} {1} (\bibinfo {year}
  {2023})}\BibitemShut {NoStop}%
\bibitem [{\citenamefont {Bhattacharya}\ \emph {et~al.}(2023)\citenamefont
  {Bhattacharya}, \citenamefont {Nandy}, \citenamefont {Nath},\ and\
  \citenamefont {Sahu}}]{bhattacharya2023krylov}%
  \BibitemOpen
  \bibfield  {author} {\bibinfo {author} {\bibfnamefont {A.}~\bibnamefont
  {Bhattacharya}}, \bibinfo {author} {\bibfnamefont {P.}~\bibnamefont {Nandy}},
  \bibinfo {author} {\bibfnamefont {P.~P.}\ \bibnamefont {Nath}},\ and\
  \bibinfo {author} {\bibfnamefont {H.}~\bibnamefont {Sahu}},\ }\bibfield
  {title} {\bibinfo {title} {On krylov complexity in open systems: an approach
  via bi-lanczos algorithm},\ }\href {https://doi.org/10.1007/JHEP12(2023)066}
  {\bibfield  {journal} {\bibinfo  {journal} {Journal of High Energy Physics}\
  }\textbf {\bibinfo {volume} {2023}},\ \bibinfo {pages} {1} (\bibinfo {year}
  {2023})}\BibitemShut {NoStop}%
\bibitem [{\citenamefont {Bhattacharjee}\ \emph {et~al.}(2024)\citenamefont
  {Bhattacharjee}, \citenamefont {Nandy},\ and\ \citenamefont
  {Pathak}}]{bhattacharjee2024operator}%
  \BibitemOpen
  \bibfield  {author} {\bibinfo {author} {\bibfnamefont {B.}~\bibnamefont
  {Bhattacharjee}}, \bibinfo {author} {\bibfnamefont {P.}~\bibnamefont
  {Nandy}},\ and\ \bibinfo {author} {\bibfnamefont {T.}~\bibnamefont
  {Pathak}},\ }\bibfield  {title} {\bibinfo {title} {Operator dynamics in
  lindbladian syk: a krylov complexity perspective},\ }\href
  {https://doi.org/10.1007/JHEP01(2024)094} {\bibfield  {journal} {\bibinfo
  {journal} {Journal of High Energy Physics}\ }\textbf {\bibinfo {volume}
  {2024}},\ \bibinfo {pages} {1} (\bibinfo {year} {2024})}\BibitemShut
  {NoStop}%
\bibitem [{\citenamefont {Liu}\ \emph {et~al.}(2023)\citenamefont {Liu},
  \citenamefont {Tang},\ and\ \citenamefont {Zhai}}]{Liu2023Krylov}%
  \BibitemOpen
  \bibfield  {author} {\bibinfo {author} {\bibfnamefont {C.}~\bibnamefont
  {Liu}}, \bibinfo {author} {\bibfnamefont {H.}~\bibnamefont {Tang}},\ and\
  \bibinfo {author} {\bibfnamefont {H.}~\bibnamefont {Zhai}},\ }\bibfield
  {title} {\bibinfo {title} {Krylov complexity in open quantum systems},\
  }\href {https://doi.org/10.1103/PhysRevResearch.5.033085} {\bibfield
  {journal} {\bibinfo  {journal} {Phys. Rev. Res.}\ }\textbf {\bibinfo {volume}
  {5}},\ \bibinfo {pages} {033085} (\bibinfo {year} {2023})}\BibitemShut
  {NoStop}%
\bibitem [{\citenamefont {Xia}\ \emph {et~al.}(2024)\citenamefont {Xia},
  \citenamefont {Zou},\ and\ \citenamefont {Li}}]{xia2024complexity}%
  \BibitemOpen
  \bibfield  {author} {\bibinfo {author} {\bibfnamefont {W.}~\bibnamefont
  {Xia}}, \bibinfo {author} {\bibfnamefont {J.}~\bibnamefont {Zou}},\ and\
  \bibinfo {author} {\bibfnamefont {X.}~\bibnamefont {Li}},\ }\href
  {https://arxiv.org/abs/2404.08055} {\bibinfo {title} {Complexity enriched
  dynamical phases for fermions on graphs}} (\bibinfo {year} {2024}),\ \Eprint
  {https://arxiv.org/abs/2404.08055} {arXiv:2404.08055 [quant-ph]} \BibitemShut
  {NoStop}%
\bibitem [{\citenamefont {Malthe-S{\o}renssen}(2024)}]{malthe2024percolation}%
  \BibitemOpen
  \bibfield  {author} {\bibinfo {author} {\bibfnamefont {A.}~\bibnamefont
  {Malthe-S{\o}renssen}},\ }\href@noop {} {\bibinfo {title} {Percolation theory
  using python}} (\bibinfo {year} {2024})\BibitemShut {NoStop}%
\bibitem [{\citenamefont {Griffiths}(1969)}]{Griffiths1969Nonanalytic}%
  \BibitemOpen
  \bibfield  {author} {\bibinfo {author} {\bibfnamefont {R.~B.}\ \bibnamefont
  {Griffiths}},\ }\bibfield  {title} {\bibinfo {title} {Nonanalytic behavior
  above the critical point in a random ising ferromagnet},\ }\href
  {https://doi.org/10.1103/PhysRevLett.23.17} {\bibfield  {journal} {\bibinfo
  {journal} {Phys. Rev. Lett.}\ }\textbf {\bibinfo {volume} {23}},\ \bibinfo
  {pages} {17} (\bibinfo {year} {1969})}\BibitemShut {NoStop}%
\bibitem [{\citenamefont {Vojta}(2010)}]{vojta2010quantum}%
  \BibitemOpen
  \bibfield  {author} {\bibinfo {author} {\bibfnamefont {T.}~\bibnamefont
  {Vojta}},\ }\bibfield  {title} {\bibinfo {title} {Quantum griffiths effects
  and smeared phase transitions in metals: theory and experiment},\ }\href
  {https://doi.org/10.1007/s10909-010-0205-4} {\bibfield  {journal} {\bibinfo
  {journal} {Journal of Low Temperature Physics}\ }\textbf {\bibinfo {volume}
  {161}},\ \bibinfo {pages} {299} (\bibinfo {year} {2010})}\BibitemShut
  {NoStop}%
\bibitem [{\citenamefont {{\v{C}}indrak}\ \emph {et~al.}(2024)\citenamefont
  {{\v{C}}indrak}, \citenamefont {Paschke}, \citenamefont {Jaurigue},\ and\
  \citenamefont {L{\"u}dge}}]{vcindrak2024measurable}%
  \BibitemOpen
  \bibfield  {author} {\bibinfo {author} {\bibfnamefont {S.}~\bibnamefont
  {{\v{C}}indrak}}, \bibinfo {author} {\bibfnamefont {A.}~\bibnamefont
  {Paschke}}, \bibinfo {author} {\bibfnamefont {L.}~\bibnamefont {Jaurigue}},\
  and\ \bibinfo {author} {\bibfnamefont {K.}~\bibnamefont {L{\"u}dge}},\
  }\bibfield  {title} {\bibinfo {title} {Measurable krylov spaces and
  eigenenergy count in quantum state dynamics},\ }\href
  {https://doi.org/10.1007/JHEP10(2024)083} {\bibfield  {journal} {\bibinfo
  {journal} {Journal of High Energy Physics}\ }\textbf {\bibinfo {volume}
  {2024}},\ \bibinfo {pages} {1} (\bibinfo {year} {2024})}\BibitemShut
  {NoStop}%
\bibitem [{\citenamefont {Meng}\ \emph {et~al.}(2020)\citenamefont {Meng},
  \citenamefont {Yu},\ and\ \citenamefont {Zhang}}]{meng2020improved}%
  \BibitemOpen
  \bibfield  {author} {\bibinfo {author} {\bibfnamefont {F.}~\bibnamefont
  {Meng}}, \bibinfo {author} {\bibfnamefont {X.}~\bibnamefont {Yu}},\ and\
  \bibinfo {author} {\bibfnamefont {Z.}~\bibnamefont {Zhang}},\ }\bibfield
  {title} {\bibinfo {title} {An improved quantum algorithm for spectral
  regression},\ }in\ \href {https://doi.org/10.1109/ACCC51160.2020.9347936}
  {\emph {\bibinfo {booktitle} {2020 Asia Conference on Computers and
  Communications (ACCC)}}}\ (\bibinfo {organization} {IEEE},\ \bibinfo {year}
  {2020})\ pp.\ \bibinfo {pages} {11--15}\BibitemShut {NoStop}%
\bibitem [{\citenamefont {Zhang}\ \emph {et~al.}(2021)\citenamefont {Zhang},
  \citenamefont {Hsieh}, \citenamefont {Liu},\ and\ \citenamefont
  {Tao}}]{Zhang2021Quantum}%
  \BibitemOpen
  \bibfield  {author} {\bibinfo {author} {\bibfnamefont {K.}~\bibnamefont
  {Zhang}}, \bibinfo {author} {\bibfnamefont {M.-H.}\ \bibnamefont {Hsieh}},
  \bibinfo {author} {\bibfnamefont {L.}~\bibnamefont {Liu}},\ and\ \bibinfo
  {author} {\bibfnamefont {D.}~\bibnamefont {Tao}},\ }\bibfield  {title}
  {\bibinfo {title} {Quantum gram-schmidt processes and their application to
  efficient state readout for quantum algorithms},\ }\href
  {https://doi.org/10.1103/PhysRevResearch.3.043095} {\bibfield  {journal}
  {\bibinfo  {journal} {Phys. Rev. Res.}\ }\textbf {\bibinfo {volume} {3}},\
  \bibinfo {pages} {043095} (\bibinfo {year} {2021})}\BibitemShut {NoStop}%
\bibitem [{\citenamefont {Li}\ and\ \citenamefont
  {xi~Liu}(2025)}]{li2025quantumalgorithmvectorset}%
  \BibitemOpen
  \bibfield  {author} {\bibinfo {author} {\bibfnamefont {Z.-M.}\ \bibnamefont
  {Li}}\ and\ \bibinfo {author} {\bibfnamefont {Y.}~\bibnamefont {xi~Liu}},\
  }\href {https://arxiv.org/abs/2412.19090} {\bibinfo {title} {Quantum
  algorithm for vector set orthogonal normalization and matrix qr decomposition
  with polynomial speedup}} (\bibinfo {year} {2025}),\ \Eprint
  {https://arxiv.org/abs/2412.19090} {arXiv:2412.19090 [quant-ph]} \BibitemShut
  {NoStop}%
\bibitem [{\citenamefont {Huang}\ \emph {et~al.}(2020)\citenamefont {Huang},
  \citenamefont {Wu}, \citenamefont {Fan},\ and\ \citenamefont
  {Zhu}}]{huang2020superconducting}%
  \BibitemOpen
  \bibfield  {author} {\bibinfo {author} {\bibfnamefont {H.-L.}\ \bibnamefont
  {Huang}}, \bibinfo {author} {\bibfnamefont {D.}~\bibnamefont {Wu}}, \bibinfo
  {author} {\bibfnamefont {D.}~\bibnamefont {Fan}},\ and\ \bibinfo {author}
  {\bibfnamefont {X.}~\bibnamefont {Zhu}},\ }\bibfield  {title} {\bibinfo
  {title} {Superconducting quantum computing: a review},\ }\href
  {https://doi.org/10.1007/s11432-020-2881-9} {\bibfield  {journal} {\bibinfo
  {journal} {Science China Information Sciences}\ }\textbf {\bibinfo {volume}
  {63}},\ \bibinfo {pages} {1} (\bibinfo {year} {2020})}\BibitemShut {NoStop}%
\bibitem [{\citenamefont {Krantz}\ \emph {et~al.}(2019)\citenamefont {Krantz},
  \citenamefont {Kjaergaard}, \citenamefont {Yan}, \citenamefont {Orlando},
  \citenamefont {Gustavsson},\ and\ \citenamefont
  {Oliver}}]{Krantz2019quantum}%
  \BibitemOpen
  \bibfield  {author} {\bibinfo {author} {\bibfnamefont {P.}~\bibnamefont
  {Krantz}}, \bibinfo {author} {\bibfnamefont {M.}~\bibnamefont {Kjaergaard}},
  \bibinfo {author} {\bibfnamefont {F.}~\bibnamefont {Yan}}, \bibinfo {author}
  {\bibfnamefont {T.~P.}\ \bibnamefont {Orlando}}, \bibinfo {author}
  {\bibfnamefont {S.}~\bibnamefont {Gustavsson}},\ and\ \bibinfo {author}
  {\bibfnamefont {W.~D.}\ \bibnamefont {Oliver}},\ }\bibfield  {title}
  {\bibinfo {title} {A quantum engineer's guide to superconducting qubits},\
  }\href {https://doi.org/10.1063/1.5089550} {\bibfield  {journal} {\bibinfo
  {journal} {Applied Physics Reviews}\ }\textbf {\bibinfo {volume} {6}},\
  \bibinfo {pages} {021318} (\bibinfo {year} {2019})},\ \Eprint
  {https://arxiv.org/abs/https://pubs.aip.org/aip/apr/article-pdf/doi/10.1063/1.5089550/16667201/021318\_1\_online.pdf}
  {https://pubs.aip.org/aip/apr/article-pdf/doi/10.1063/1.5089550/16667201/021318\_1\_online.pdf}
  \BibitemShut {NoStop}%
\bibitem [{\citenamefont {Kjaergaard}\ \emph {et~al.}(2020)\citenamefont
  {Kjaergaard}, \citenamefont {Schwartz}, \citenamefont {Braumüller},
  \citenamefont {Krantz}, \citenamefont {Wang}, \citenamefont {Gustavsson},\
  and\ \citenamefont {Oliver}}]{Kjaergaard2020Superconducting}%
  \BibitemOpen
  \bibfield  {author} {\bibinfo {author} {\bibfnamefont {M.}~\bibnamefont
  {Kjaergaard}}, \bibinfo {author} {\bibfnamefont {M.~E.}\ \bibnamefont
  {Schwartz}}, \bibinfo {author} {\bibfnamefont {J.}~\bibnamefont
  {Braumüller}}, \bibinfo {author} {\bibfnamefont {P.}~\bibnamefont {Krantz}},
  \bibinfo {author} {\bibfnamefont {J.~I.-J.}\ \bibnamefont {Wang}}, \bibinfo
  {author} {\bibfnamefont {S.}~\bibnamefont {Gustavsson}},\ and\ \bibinfo
  {author} {\bibfnamefont {W.~D.}\ \bibnamefont {Oliver}},\ }\bibfield  {title}
  {\bibinfo {title} {Superconducting qubits: Current state of play},\ }\href
  {https://doi.org/https://doi.org/10.1146/annurev-conmatphys-031119-050605}
  {\bibfield  {journal} {\bibinfo  {journal} {Annual Review of Condensed Matter
  Physics}\ }\textbf {\bibinfo {volume} {11}},\ \bibinfo {pages} {369}
  (\bibinfo {year} {2020})}\BibitemShut {NoStop}%
\bibitem [{\citenamefont {Wendin}(2017)}]{Wendin2017Quantum}%
  \BibitemOpen
  \bibfield  {author} {\bibinfo {author} {\bibfnamefont {G.}~\bibnamefont
  {Wendin}},\ }\bibfield  {title} {\bibinfo {title} {Quantum information
  processing with superconducting circuits: a review},\ }\href
  {https://doi.org/10.1088/1361-6633/aa7e1a} {\bibfield  {journal} {\bibinfo
  {journal} {Reports on Progress in Physics}\ }\textbf {\bibinfo {volume}
  {80}},\ \bibinfo {pages} {106001} (\bibinfo {year} {2017})}\BibitemShut
  {NoStop}%
\bibitem [{\citenamefont {Wu}\ \emph {et~al.}(2021)\citenamefont {Wu},
  \citenamefont {Liang}, \citenamefont {Tian}, \citenamefont {Yang},
  \citenamefont {Chen}, \citenamefont {Liu}, \citenamefont {Tey},\ and\
  \citenamefont {You}}]{Wu2021concise}%
  \BibitemOpen
  \bibfield  {author} {\bibinfo {author} {\bibfnamefont {X.}~\bibnamefont
  {Wu}}, \bibinfo {author} {\bibfnamefont {X.}~\bibnamefont {Liang}}, \bibinfo
  {author} {\bibfnamefont {Y.}~\bibnamefont {Tian}}, \bibinfo {author}
  {\bibfnamefont {F.}~\bibnamefont {Yang}}, \bibinfo {author} {\bibfnamefont
  {C.}~\bibnamefont {Chen}}, \bibinfo {author} {\bibfnamefont {Y.-C.}\
  \bibnamefont {Liu}}, \bibinfo {author} {\bibfnamefont {M.~K.}\ \bibnamefont
  {Tey}},\ and\ \bibinfo {author} {\bibfnamefont {L.}~\bibnamefont {You}},\
  }\bibfield  {title} {\bibinfo {title} {A concise review of rydberg atom based
  quantum computation and quantum simulation*},\ }\href
  {https://doi.org/10.1088/1674-1056/abd76f} {\bibfield  {journal} {\bibinfo
  {journal} {Chinese Physics B}\ }\textbf {\bibinfo {volume} {30}},\ \bibinfo
  {pages} {020305} (\bibinfo {year} {2021})}\BibitemShut {NoStop}%
\bibitem [{\citenamefont {Adams}\ \emph {et~al.}(2019)\citenamefont {Adams},
  \citenamefont {Pritchard},\ and\ \citenamefont {Shaffer}}]{Adams2020Rydberg}%
  \BibitemOpen
  \bibfield  {author} {\bibinfo {author} {\bibfnamefont {C.~S.}\ \bibnamefont
  {Adams}}, \bibinfo {author} {\bibfnamefont {J.~D.}\ \bibnamefont
  {Pritchard}},\ and\ \bibinfo {author} {\bibfnamefont {J.~P.}\ \bibnamefont
  {Shaffer}},\ }\bibfield  {title} {\bibinfo {title} {Rydberg atom quantum
  technologies},\ }\href {https://doi.org/10.1088/1361-6455/ab52ef} {\bibfield
  {journal} {\bibinfo  {journal} {Journal of Physics B: Atomic, Molecular and
  Optical Physics}\ }\textbf {\bibinfo {volume} {53}},\ \bibinfo {pages}
  {012002} (\bibinfo {year} {2019})}\BibitemShut {NoStop}%
\bibitem [{\citenamefont {Saffman}\ \emph {et~al.}(2010)\citenamefont
  {Saffman}, \citenamefont {Walker},\ and\ \citenamefont
  {M\o{}lmer}}]{Saffman2010Quantum}%
  \BibitemOpen
  \bibfield  {author} {\bibinfo {author} {\bibfnamefont {M.}~\bibnamefont
  {Saffman}}, \bibinfo {author} {\bibfnamefont {T.~G.}\ \bibnamefont
  {Walker}},\ and\ \bibinfo {author} {\bibfnamefont {K.}~\bibnamefont
  {M\o{}lmer}},\ }\bibfield  {title} {\bibinfo {title} {Quantum information
  with rydberg atoms},\ }\href {https://doi.org/10.1103/RevModPhys.82.2313}
  {\bibfield  {journal} {\bibinfo  {journal} {Rev. Mod. Phys.}\ }\textbf
  {\bibinfo {volume} {82}},\ \bibinfo {pages} {2313} (\bibinfo {year}
  {2010})}\BibitemShut {NoStop}%
\bibitem [{\citenamefont {Saffman}(2016)}]{Saffman2016Quantumcomputing}%
  \BibitemOpen
  \bibfield  {author} {\bibinfo {author} {\bibfnamefont {M.}~\bibnamefont
  {Saffman}},\ }\bibfield  {title} {\bibinfo {title} {Quantum computing with
  atomic qubits and rydberg interactions: progress and challenges},\ }\href
  {https://doi.org/10.1088/0953-4075/49/20/202001} {\bibfield  {journal}
  {\bibinfo  {journal} {Journal of Physics B: Atomic, Molecular and Optical
  Physics}\ }\textbf {\bibinfo {volume} {49}},\ \bibinfo {pages} {202001}
  (\bibinfo {year} {2016})}\BibitemShut {NoStop}%
\bibitem [{\citenamefont {Häffner}\ \emph {et~al.}(2008)\citenamefont
  {Häffner}, \citenamefont {Roos},\ and\ \citenamefont
  {Blatt}}]{Haffner2008Quantum}%
  \BibitemOpen
  \bibfield  {author} {\bibinfo {author} {\bibfnamefont {H.}~\bibnamefont
  {Häffner}}, \bibinfo {author} {\bibfnamefont {C.}~\bibnamefont {Roos}},\
  and\ \bibinfo {author} {\bibfnamefont {R.}~\bibnamefont {Blatt}},\ }\bibfield
   {title} {\bibinfo {title} {Quantum computing with trapped ions},\ }\href
  {https://doi.org/https://doi.org/10.1016/j.physrep.2008.09.003} {\bibfield
  {journal} {\bibinfo  {journal} {Physics Reports}\ }\textbf {\bibinfo {volume}
  {469}},\ \bibinfo {pages} {155} (\bibinfo {year} {2008})}\BibitemShut
  {NoStop}%
\bibitem [{\citenamefont {Bruzewicz}\ \emph {et~al.}(2019)\citenamefont
  {Bruzewicz}, \citenamefont {Chiaverini}, \citenamefont {McConnell},\ and\
  \citenamefont {Sage}}]{Bruzewicz2019Trapped}%
  \BibitemOpen
  \bibfield  {author} {\bibinfo {author} {\bibfnamefont {C.~D.}\ \bibnamefont
  {Bruzewicz}}, \bibinfo {author} {\bibfnamefont {J.}~\bibnamefont
  {Chiaverini}}, \bibinfo {author} {\bibfnamefont {R.}~\bibnamefont
  {McConnell}},\ and\ \bibinfo {author} {\bibfnamefont {J.~M.}\ \bibnamefont
  {Sage}},\ }\bibfield  {title} {\bibinfo {title} {Trapped-ion quantum
  computing: Progress and challenges},\ }\href
  {https://doi.org/10.1063/1.5088164} {\bibfield  {journal} {\bibinfo
  {journal} {Applied Physics Reviews}\ }\textbf {\bibinfo {volume} {6}},\
  \bibinfo {pages} {021314} (\bibinfo {year} {2019})},\ \Eprint
  {https://arxiv.org/abs/https://pubs.aip.org/aip/apr/article-pdf/doi/10.1063/1.5088164/19742554/021314\_1\_online.pdf}
  {https://pubs.aip.org/aip/apr/article-pdf/doi/10.1063/1.5088164/19742554/021314\_1\_online.pdf}
  \BibitemShut {NoStop}%
\bibitem [{\citenamefont {Blatt}\ and\ \citenamefont
  {Roos}(2012)}]{blatt2012quantum}%
  \BibitemOpen
  \bibfield  {author} {\bibinfo {author} {\bibfnamefont {R.}~\bibnamefont
  {Blatt}}\ and\ \bibinfo {author} {\bibfnamefont {C.~F.}\ \bibnamefont
  {Roos}},\ }\bibfield  {title} {\bibinfo {title} {Quantum simulations with
  trapped ions},\ }\href {https://doi.org/10.1038/nphys2252} {\bibfield
  {journal} {\bibinfo  {journal} {Nature Physics}\ }\textbf {\bibinfo {volume}
  {8}},\ \bibinfo {pages} {277} (\bibinfo {year} {2012})}\BibitemShut {NoStop}%
\bibitem [{\citenamefont {Monroe}\ \emph {et~al.}(2021)\citenamefont {Monroe},
  \citenamefont {Campbell}, \citenamefont {Duan}, \citenamefont {Gong},
  \citenamefont {Gorshkov}, \citenamefont {Hess}, \citenamefont {Islam},
  \citenamefont {Kim}, \citenamefont {Linke}, \citenamefont {Pagano},
  \citenamefont {Richerme}, \citenamefont {Senko},\ and\ \citenamefont
  {Yao}}]{Monroe2021Programmable}%
  \BibitemOpen
  \bibfield  {author} {\bibinfo {author} {\bibfnamefont {C.}~\bibnamefont
  {Monroe}}, \bibinfo {author} {\bibfnamefont {W.~C.}\ \bibnamefont
  {Campbell}}, \bibinfo {author} {\bibfnamefont {L.-M.}\ \bibnamefont {Duan}},
  \bibinfo {author} {\bibfnamefont {Z.-X.}\ \bibnamefont {Gong}}, \bibinfo
  {author} {\bibfnamefont {A.~V.}\ \bibnamefont {Gorshkov}}, \bibinfo {author}
  {\bibfnamefont {P.~W.}\ \bibnamefont {Hess}}, \bibinfo {author}
  {\bibfnamefont {R.}~\bibnamefont {Islam}}, \bibinfo {author} {\bibfnamefont
  {K.}~\bibnamefont {Kim}}, \bibinfo {author} {\bibfnamefont {N.~M.}\
  \bibnamefont {Linke}}, \bibinfo {author} {\bibfnamefont {G.}~\bibnamefont
  {Pagano}}, \bibinfo {author} {\bibfnamefont {P.}~\bibnamefont {Richerme}},
  \bibinfo {author} {\bibfnamefont {C.}~\bibnamefont {Senko}},\ and\ \bibinfo
  {author} {\bibfnamefont {N.~Y.}\ \bibnamefont {Yao}},\ }\bibfield  {title}
  {\bibinfo {title} {Programmable quantum simulations of spin systems with
  trapped ions},\ }\href {https://doi.org/10.1103/RevModPhys.93.025001}
  {\bibfield  {journal} {\bibinfo  {journal} {Rev. Mod. Phys.}\ }\textbf
  {\bibinfo {volume} {93}},\ \bibinfo {pages} {025001} (\bibinfo {year}
  {2021})}\BibitemShut {NoStop}%
\bibitem [{\citenamefont {Duan}\ and\ \citenamefont
  {Monroe}(2010)}]{Duan2010Colloquium}%
  \BibitemOpen
  \bibfield  {author} {\bibinfo {author} {\bibfnamefont {L.-M.}\ \bibnamefont
  {Duan}}\ and\ \bibinfo {author} {\bibfnamefont {C.}~\bibnamefont {Monroe}},\
  }\bibfield  {title} {\bibinfo {title} {Colloquium: Quantum networks with
  trapped ions},\ }\href {https://doi.org/10.1103/RevModPhys.82.1209}
  {\bibfield  {journal} {\bibinfo  {journal} {Rev. Mod. Phys.}\ }\textbf
  {\bibinfo {volume} {82}},\ \bibinfo {pages} {1209} (\bibinfo {year}
  {2010})}\BibitemShut {NoStop}%
\bibitem [{\citenamefont {Monroe}\ and\ \citenamefont
  {Kim}(2013)}]{Monroe2013Scaling}%
  \BibitemOpen
  \bibfield  {author} {\bibinfo {author} {\bibfnamefont {C.}~\bibnamefont
  {Monroe}}\ and\ \bibinfo {author} {\bibfnamefont {J.}~\bibnamefont {Kim}},\
  }\bibfield  {title} {\bibinfo {title} {Scaling the ion trap quantum
  processor},\ }\href {https://doi.org/10.1126/science.1231298} {\bibfield
  {journal} {\bibinfo  {journal} {Science}\ }\textbf {\bibinfo {volume}
  {339}},\ \bibinfo {pages} {1164} (\bibinfo {year} {2013})},\ \Eprint
  {https://arxiv.org/abs/https://www.science.org/doi/pdf/10.1126/science.1231298}
  {https://www.science.org/doi/pdf/10.1126/science.1231298} \BibitemShut
  {NoStop}%
\end{thebibliography}

%

\clearpage

\begin{widetext} 


\renewcommand{\thesection}{S\arabic{section}}      
\renewcommand{\thesubsection}{S\arabic{section}-\arabic{subsection}} 
\renewcommand{\theequation}{S\arabic{equation}}     
\renewcommand{\thefigure}{S\arabic{figure}}         
\renewcommand{\thetable}{S\arabic{table}}           

\setcounter{section}{0}    
\setcounter{equation}{0}
\setcounter{figure}{0}
\setcounter{table}{0}

\newpage

\begin{center} 
{\Huge \bf Supplementary Materials} \\
\end{center} 
The Supplementary materials primarily include: foundational concepts from percolation theory; detailed calculations of Krylov complexity in one-dimensional systems; numerical results on Bethe lattices, two-dimensional lattices, and Penrose lattices; finite-size scaling analyses; and comprehensive details of the proposed experimental implementations.

\tableofcontents

\newpage

\section{    Classical site percolation theory} 
In this section, we provide a brief introduction to classical percolation theory and use it to evaluate the Krylov dimension. We begin by defining the basic concepts and terms.
\begin{definition}

\hspace*{\fill}

\begin{enumerate}
    \item Two sites are \textbf{connected} if they are nearest neighbors.
    \item A \textbf{cluster} is a set of connected sites.
    \item A cluster is \textbf{spanning} if it spans from one side to the opposite side.
    \item A cluster that is spanning is called the \textbf{spanning cluster}.
    \item A system is \textbf{percolating} if there is a spanning cluster in the system.
\end{enumerate}
\end{definition}

\begin{definition}
Percolation probability $\Pi(p, L)$

\hspace*{\fill}

The percolation probability $\Pi(p, L)$ is the probability for there to be a connected path from one side to another side as a function of $p$ in a system of size $L$.
\end{definition}

\begin{definition}
Density of the spanning cluster $P(p, L)$

\hspace*{\fill}

The probability $P(p, L)$ for a site to belong to a spanning cluster is called the density of the spanning cluster, or the order parameter for the percolation problem.
\end{definition}

\subsection{One-dimension percolation}

\subsubsection{Percolation probability $\Pi(p, L)$}
In this case, there is a spanning cluster if and only if all the sites are occupied. Therefore, the percolation probability is
\begin{equation}
    \Pi(p, L) = p^L.
\end{equation}
In the thermal limit $L\rightarrow \infty$,
\begin{equation}
    \Pi(p, \infty) = 
\begin{cases}
0, & p<1,\\
1, & p = 1 .
\end{cases}
\end{equation}
Thus, the percolation threshold in one dimension is $p_c = 1$. 
\subsubsection{Cluster number density $n(s, p)$}
The probability of a randomly selected site in the lattice being part of a cluster with size $s$ is represented by:
\begin{equation}
    P(\text{site is part of a cluster of size }s) = s\rho(s,p).
\end{equation}
Here, $\rho(s,p)$ denotes the cluster number density, indicating the probability of a site belonging to a specific site within a cluster of size $s$. To form a cluster of size s, the connected s sizes must be present, and the right and left ends must be absent. Thus, the $\rho(s,p)$ is
\begin{equation}
    \rho(s,p) = (1 - p)^2p^s.
\end{equation} 
\subsubsection{Average cluster size $S(p)$}
The average cluster size $S$ is defined as 
\begin{eqnarray}
    S(p) = \braket{s} & = & \sum_{s = 1} s \frac{s\rho(s,p)}{\sum_s s\rho(s,p)} = \frac{1}{p}  \sum_{s = 1} s^2\rho(s,p) \\
    & = &\frac{(1-p)^2}{p}\sum_{s = 1} s^2 p^s \\
    & = & \frac{(1-p)^2}{p} \sum_{s = 1} p\frac{\partial }{\partial p} p \frac{\partial }{\partial p} p^s \\
     & = & \frac{1 + p}{1 - p}
\end{eqnarray}
Here, $\sum_{s = 1}^{\infty} s\rho(s,p)  = (1 - p)^2p(1-p)^{-2} = p$ and we introduce the critical exponent $\gamma = 1$,
\begin{equation}
    S = \frac{1 + p}{1 - p} = \frac{\Gamma(p)}{|p - p_c|^{\gamma}},
\end{equation}
where $\Gamma(p) = 1 + p$.

\subsection{Percolation on Bette lattice}
The Bette lattice is a tree structure in which each node has $Z$ neighbors. This structure has no loops and we need an infinite number of dimensions in Euclidean space to embed a tree without loops. Thus, the Bette lattice is an infinite dimension. 
\subsubsection{Percolation threshold $p_c$}
To get a spanning cluster, we need to ensure that at least one of the neighbor $Z-1$ sites is occupied on average. Thus, 
\begin{equation}
    p_c = \frac{1}{Z - 1}
\end{equation}
\subsubsection{Average cluster size $S(p)$}
We define $T(p)$ as the average number of sites connected to a given site on a specific branch. Thus, 
\begin{equation}
    S = 1 + ZT.
\end{equation}
Here, $1$ represents the central point. To investigate $T(p)$, we should consider the following cases. Starting from the central site, the next site $k$ is empty with probability $1 - p$, yielding a contribution of $0$. Alternatively, if the next site $k$ is occupied with probability $p$, the contribution is $1$ (from the site $k$) plus $Z-1$ sub-branches. Thus, we have:
\begin{eqnarray}
    T &=& (1-p)\times 0 + p(1 + (Z-1)T),\\
    T &=&  \frac{p}{1 - p(Z -1)}.
\end{eqnarray}
The average cluster size $S(p)$ is given by:
\begin{equation}
    S(p) = 1 + ZT = \frac{1 + p}{1 - (Z -1)p} = \frac{p_c(1+p)}{p_c - p} = \frac{\Gamma}{|p_c - p|^{\gamma}},
\end{equation}
where $p_c = \frac{1}{Z -1}$. Here, $\Gamma(p) = p_c(1 + p)$ and $\gamma = 1$.

\subsection{Scaling theory for $\rho(s, p)$}
The general form of the cluster density is 
\begin{equation}
    \rho(s,p) = s^{-\tau}F(s/s_{\xi}) = s^{-\tau}F((p-p_c)s^{\sigma}),
\end{equation}
where $s_{\xi} = s_0|p-p_c|^{-1/\sigma}$ is the correlation length. An alternative scaling form is
\begin{equation}
    \rho(s,p) = s^{-\tau}\hat{F}((p-p_c)s^{\sigma}),
\end{equation}
where $\hat{F}(u) = F(u^{\sigma})$. The scaling function  $\hat{F}(u)$ goes exponentially
fast to zero. This simple scaling ansatz leads to a scaling relation between critical exponents. The average cluster size is
\begin{equation}
    S(p) = \sum_s s^2\rho(s,p) = \int s^2\rho(s,p)ds = \int_1^{\infty} s^{2-\tau}\hat{F}((p-p_c)s^{\sigma})ds.
\end{equation}
And the result for the average cluster size is
\begin{equation}
    S(p) \sim (p_c-p)^{\frac{\tau-3}{\sigma}} \sim \frac{1}{(p_c-p)^{\gamma}} 
\end{equation}
which gives a scaling relation $\gamma = \frac{3-\tau}{\sigma}$.

\section{Quantum Krylov complexity phase transition}

\subsection{Average Krylov dimension of one-dimension quantum percolation model}
Building upon our previous work, we establish a reasonable assumption that the Krylov dimension of a free Hamiltonian is proportional to the square of the number of sites connected to the site where the initial operator is located. The dimension $D_{\text{Free}}$ of the Krylov space is determined as follows:
\begin{eqnarray}
    D_{\text{Free}} &=& \sum_{s = 1}^{\infty} s^2\times s\rho(s,p),\\
    D_{\text{Free}} &=& \frac{p}{(1-p)^2}(p^2+4p+1), \\
    D_{\text{Free}} &=& \frac{\Delta(p)}{|p- p_c|^{\delta}},
\end{eqnarray}
where $\Delta(p) = p(p^2+4p+1)$ and the critical exponent $\delta = 2$. The Krylov dimension diverges at the critical probability $p_c = 1$, consistent with the percolation transitions. 

For interacting fermions, the Krylov dimension is proportional to the cluster size $s$, represented as $\sim 4^{s}$. Hence, the dimension $D_{\text{Int}}$ of the Krylov space for interacting fermions is determined as follows:
\begin{eqnarray}
    D_{\text{Int}} &=& \sum_{s = 1}^{\infty} 4^s\times s\rho(s,p),\\
    D_{\text{Int}} &=& (1 - p)^2p\frac{\partial}{\partial p}\sum_{s =1}^{\infty}(4p)^s, \\
    D_{\text{Int}} &=& (1 - p)^2p\frac{\partial}{\partial p}\frac{1}{1-4p}, \\
    D_{\text{Int}} &=& \frac{4p(1-p)^2}{(1-4p)^2}.
\end{eqnarray}
Interestingly, the critical probability $p_c$ is $1/4$, which is not the same as percolation transitions.

\subsection{Finite size effect}
In the theoretical calculation of the main text, we focus only on the thermal dynamical limit $L\rightarrow \infty$. Here,  we consider the finite size effect. First, as a simple example, we investigate the one-dimensional lattice without interaction. 
In the main text, our theoretical analysis is restricted to the thermodynamic limit 
$L\rightarrow \infty$. Here, we examine finite-size effects. As a simple example, we consider a one-dimensional, non-interacting lattice. The corresponding expression for $D_{1D}^{\text{Free}}$ is given by:

\begin{align*}
 D_{1D}^{\text{Free}} &= \sum_{s = 1}^{L} s^2 s\rho(s,p), \\
                      &= (1-p)^2\sum_{s = 1}^{L} s^3p^s, \\
                      &=(1-p)^2p\frac{\partial}{\partial p}p\frac{\partial}{\partial p}p\frac{\partial}{\partial p}(\sum_{s = 1}^{L} p^s),\\
                      &=(1-p)^2p\frac{\partial}{\partial p}p\frac{\partial}{\partial p}p\frac{\partial}{\partial p}(\frac{1-p^L}{1-p}).
\end{align*}
We focus on the leading-order behavior in $L$, which scales as $L^3(1- p)p^{L}$. This result indicates that $D_{1D}^{\text{Free}}$ diverges as $L^3$ in the large-size limit. A similar calculation can be performed for the one-dimensional lattice with interactions. In this case, the corresponding leading-order term is $\frac{L^2(1 - p^2)(4p)^L}{(1 - 4p)}$. This implies that $ D_{1D}^{\text{Int}} $ diverges as $ L^2 $ in the large-size limit.

Having analyzed the one-dimensional case, we now proceed to a more general scaling analysis. We introduce the following scaling ansatz:
\begin{equation}
    D(p, L) = L^X \Sigma\left(\frac{L}{\xi}\right),
\end{equation}
where $ X $ is an unknown critical exponent, and $ \xi $ denotes the correlation length, which scales as $ \xi = \xi_0 |p - p_c|^{-\nu} $.

In the regime $ L \ll \xi $, the function $ D(p, L) $ should depend only on the system size $L$, implying that $ \Sigma(u) $ approaches a constant as $ u \ll 1$. Conversely, in the regime $ L \gg \xi $, $ D(p, L) $ diverges as $ p \to p_c $. In this limit, we expect:
\begin{equation}
    D \propto |p - p_c|^{-\gamma} \propto \xi^{\gamma/\nu}.
\end{equation}

For $ D(p, L) $ to become independent of $ L $ in this limit, the scaling function $ \Sigma(u) $ must scale as a power law with exponent $ -X $, i.e., $ \Sigma(u) \propto u^{-X} $ when $ u \gg 1 $. Substituting this into the scaling ansatz gives:
\begin{equation}
    D(p, L) = L^X \Sigma\left(\frac{L}{\xi}\right) \propto L^X \left(\frac{L}{\xi}\right)^{-X} \propto \xi^X.
\end{equation}

Comparing this with the earlier expression $ D \propto \xi^{\gamma/\nu} $, we find $ X = \gamma/\nu $. Therefore, the full finite-size scaling form is:
\begin{equation}
    D(p, L) = L^{\gamma/\nu} \Sigma\left(\frac{L}{\xi}\right),
\end{equation}
where the scaling function \( \Sigma(u) \) behaves as
\begin{equation}
    \Sigma(u) = 
    \begin{cases}
        \text{constant}, & u \ll 1, \\
        u^{-\gamma/\nu}, & u \gg 1.
    \end{cases}
\end{equation}

\subsection{Percolation on Bethe lattice, two-dimensional lattice, and Penrose lattice}
We consider the quantum percolation model on the Bethe lattice. The Bethe lattice is a tree structure where each node has $ Z $ neighbors. To achieve percolation, at least one of the $ Z-1 $ neighboring sites must be occupied on average. Thus, $ p_c = \frac{1}{Z - 1} $. The cluster number density $ n(s,p) $ on the Bethe lattice is given by:
\begin{equation}
    n(s,p) = Cs^{-5/2}e^{-s/s_{\xi}},
\end{equation}
where $ s_{\xi} $ is the characteristic cluster size, defined as $ s_{\xi} = B^{-1}(p - p_c)^{-2} $. Here, $ B $ and $ C $ are constants. For the free quantum percolation model defined on the Bethe lattice, the Krylov dimension is:
\begin{equation}
    D_{\text{Free}} = \sum_{s = 0}^{\infty} s^3 \times Cs^{-5/2}e^{-s/s_{\xi}} = \frac{C'}{(p - p_c)^{\delta}},
\end{equation}
where $ C' $ is a constant and $ \delta = 3 $. We also find the critical probability $ p_{c,k} $ is equal to $ p_c $ and different critical exponent. The average cluster size $\braket{S}$ for classical percolation on Bette lattice is $\frac{p_c(1+p)}{p_c - p}$ and the critical exponent $\gamma = 1$. 

The other case we consider is free quantum percolation on the two-dimensional square lattice. In this case, there is no exact solution, and the critical probability $p_c = 0.6$ is obtained numerically. We also investigate the Krylov dimension, assuming that the Krylov dimension of a cluster of size $s$ is $s^2$. We analyze an $L \times L$ square lattice, with results shown in Fig.~\ref{fig:TwoDimensionFree}. The Krylov dimension increases sharply from $p = 0.55$ to $p = 0.6$. The inset displays the derivative $\frac{\partial \log(D)}{\partial p}$, which shows a significant peak at $p = 0.6$. The black dotted line represents the critical probability of classical percolation. For free quantum percolation on the two-dimensional square lattice, we also find that $p_c^{k} = p_c$. According to the scaling theory for $n(s, p)$ in classical site percolation, we have $n(s,p) = s^{-\tau} \hat{F}((p_c - p) s^{\sigma})$, where $\hat{F} = F(u^\sigma)$ decays exponentially fast to zero. The critical exponents $\tau = \frac{187}{91}$ and $\sigma = \frac{36}{91}$ are specific to classical site percolation on the two-dimensional lattice. Thus, the Krylov dimension is given by:
\begin{equation}
    \sum_{s = 0}^{\infty} s^3 \times s^{-\tau} \hat{F}((p_c - p) s^{\sigma}) = \Gamma (p_c - p)^{\frac{\tau - 4}{\sigma}},
\end{equation}
where $\Gamma$ is a constant. The critical exponent of quantum site percolation is $\frac{4 - \tau}{\sigma} = \frac{177}{36}$, while the critical exponent $\gamma$ for classical percolation on the two-dimensional lattice is $\frac{3 - \tau}{\sigma} = \frac{43}{18}$.

\begin{figure}
    \vspace{12pt}
    \centering
    \includegraphics[width=0.5\linewidth]{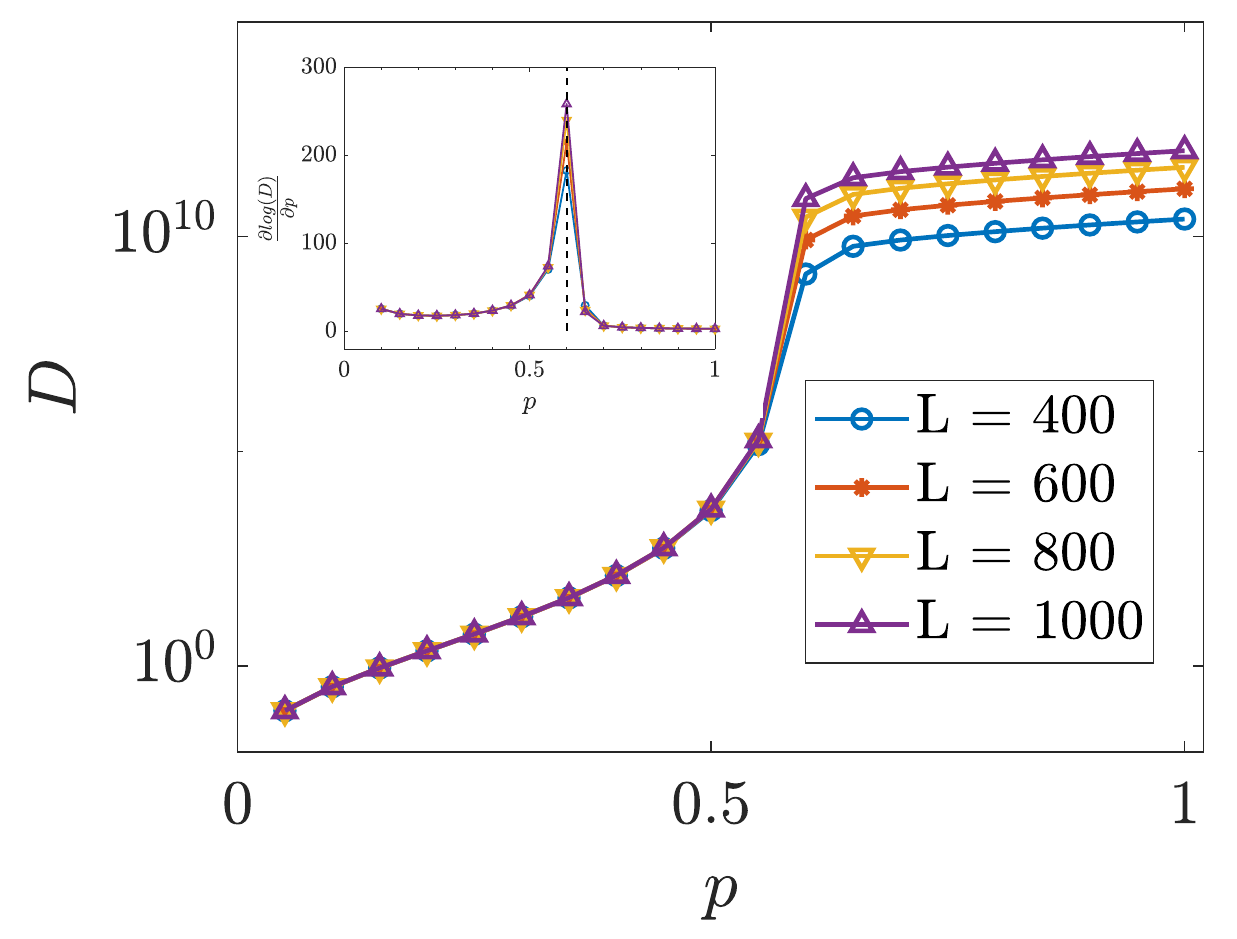}
    \caption{The Krylov dimension $D$ of two-dimensional percolation. We investigate the Krylov dimension of the quantum percolation model on a two-dimensional lattice. We consider different system sizes $L \times L$, where $L = 400, 600, 800, 1000$, and sample 1000 configurations and initial operators for each $p$. The inset shows the derivative of $D$, $\frac{\partial \log(D)}{\partial p}$. The black dotted line represents the classical percolation transition, which matches the peak of the derivative.}
    \label{fig:TwoDimensionFree}
\end{figure}

The scaling theory applies not just to the periodic lattices mentioned above, but also to certain aperiodic lattices like quasicrystals. Percolation transition observed in various quasicrystals has been found to adhere to this scaling theory. Here, we illustrate a representative two-dimensional quasicrystal, Penrose tiling. The critical exponents of a two-dimensional Penrose tiling are identical to those of a two-dimensional lattice, indicating their classification within a universal group. Penrose tiling is a well-known example of a two-dimensional quasicrystal, characterized by non-periodic tiling, with various forms of this model existing. Here, we focus on the P3 Penrose tiling, which consists of a thick rhombus with an acute angle of 72° and a thin rhombus with an acute angle of 36°, as shown in Fig.~\ref{fig:Penrose} (a).

We investigate the free quantum percolation model on this lattice, assuming that the Krylov dimension of a cluster with size $s$ is $s^2$. The corresponding results are presented in Fig.~\ref{fig:Penrose} (b). The system size is 4181. The inset shows the derivative of the Krylov dimension, with the black dotted line indicating the critical value $p_c = 0.58$, which coincides with the percolation threshold.
\begin{figure}
    \centering
    \includegraphics[width=1\linewidth]{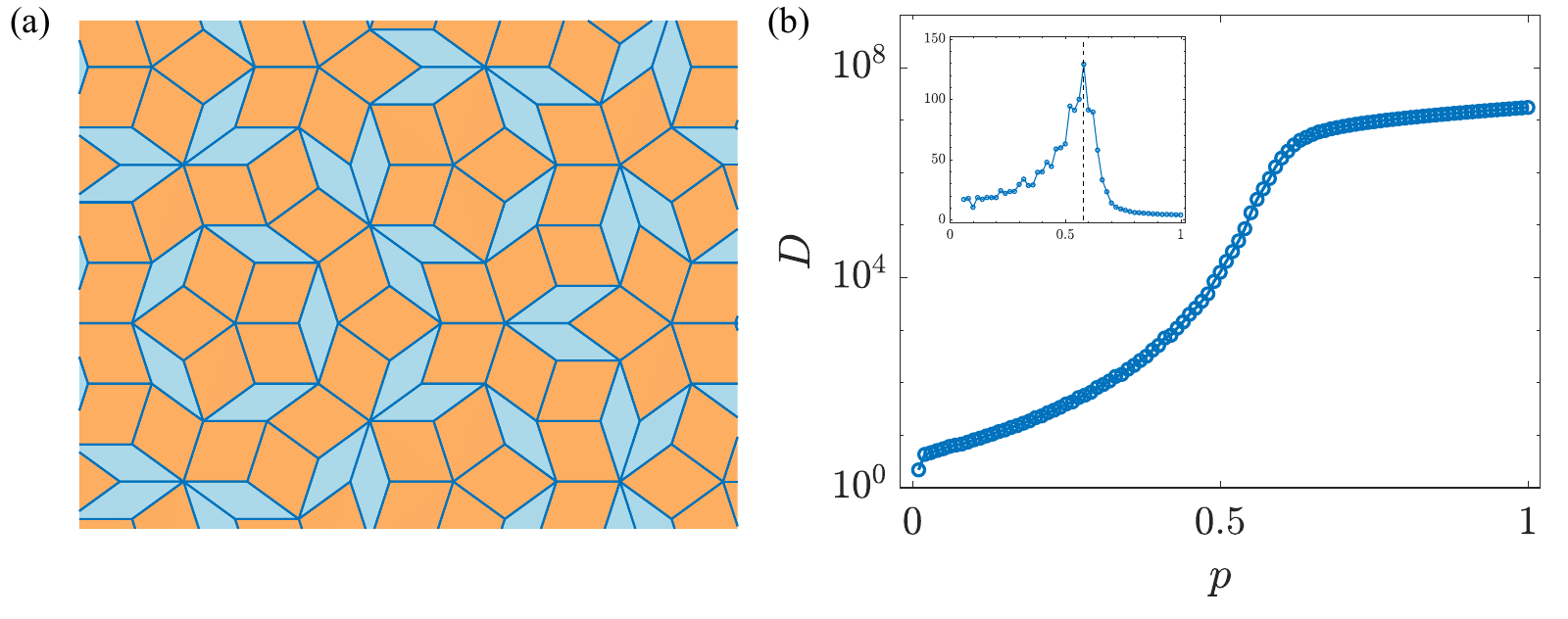}
    \caption{The results for the Penrose lattice. (a) a typical diagram of P3 Penrose tiling. (b) the Krylov dimension of the free quantum percolation model on the Penrose lattice. The inset shows the derivative of the Krylov dimension $D$ with respect to the probability $p$. The black dotted line represents the critical probability $p_c = 0.58$.}
    \label{fig:Penrose}
\end{figure}

We have investigated the Krylov dimension of the free quantum percolation model on one-dimensional, two-dimensional, Bethe lattices and Penrose tiling. We find that the critical probabilities for these lattices match those of classical percolation. This result is interpretable. In our model, sites are present with probability $p$, and the system is divided into clusters. The Hamiltonian is defined on these clusters, and the corresponding Krylov dimension depends on the cluster size. Since the Hamiltonians of different clusters commute, as $p$ is continuously increased, the system suddenly generates spanning clusters around the critical probability. This sudden formation of spanning clusters leads to a sharp increase in the Krylov dimension above the critical probability. Therefore, we believe that a Krylov complexity phase transition will also occur when considering other percolation variants, such as explosive percolation and shortest-path percolation.

Quantitative understanding can be obtained from scaling theory. The above lattices all obey the scaling theory of classical percolation $n(s,p) = s^{-\tau} \hat{F}((p_c - p) s^{\sigma})$. With the $n(s,p)$, we have established the scaling law for the Krylov dimension as $D = \Gamma (p_c - p)^{\frac{\tau - 4}{\sigma}}$, which differs from the average cluster size $\braket{S} = \Gamma (p_c - p)^{\frac{\tau - 3}{\sigma}}$ in classical percolation. Thus, the Krylov complexity phase transition on the free quantum percolation model has a corresponding relationship with the classical percolation, sharing the same critical threshold and having different scaling laws.  

\section{Experimental proposal}
\subsection{The measurable Krylov spaces}
In the main text, we introduce the conception of Krylov space based on the Heisenberg picture. The Krylov space of the operator is spanned by $K_n = \{\mathcal{L}^n|O)\}$. Similarly, the Krylov space of the state also can be defined. The time evolution is 
\begin{equation}
    \ket{\psi (t)} = e^{-iHt}\ket{\psi_0} = \sum_n \frac{t^n}{n!}(-iH)^n\ket{\psi_0}.
\end{equation}
$\ket{\psi_0}$ is the initial state. Thus, the Krylov space of the state is spanned by $(-iH)^n\ket{\psi_0}$. In the referece~\cite{vcindrak2024measurable},  the Krylov space of state can be constructed from different time-evolved states of the system, and the constructed space is measurable. Here, we generalize the formalism to the Krylov space of the operator. With discrete times $0 = t_0<t_1<\cdots<t_{n-1} = T_P$ and $T_P$ is the period of the time evolution, we introduce the operator vector
\begin{equation}
    h_i^{(n)}(|O_0)) := \sum_{j = 0}^{n - 1}f^j(|O_0))\frac{t^j_i}{j!},
\end{equation}
where $f^j(|O_0)) := (i\mathcal{L})^j|O_0)$. The vectors $h_i^{(n)}(|O_0))$ span a space $H_n^n$,
\begin{equation}
    H_n^n = \text{Span}(h_0^{(n)}(|O_0)),h_1^{(n)}(|O_0)),\cdots,h_(n-1)^{(n)}(|O_0))). 
\end{equation}
In the reference~\cite{vcindrak2024measurable}, Theorem 2 has proved $H_n^n = K_n$. For every discrete time $t_i$, we introduce the sampling operator vectors of the time evolution
\begin{equation}
    |g_i) = |O(t_i)) = e^{i\mathcal{L}t_i}|O_0).
\end{equation}
Then for a large $N_L$, we define 
\begin{equation}
    G_{N_L} = \text{Span}(|g_0),\cdots,|g_{N_L-1})).
\end{equation}
In the reference~\cite{vcindrak2024measurable}, the Theorem 3 has proved $H_{N_L}^{N_L}\approx G_{N_L}$. From this it follows then that $G_{N_L}\approx K_{N_L}$ holds as well and finally $G_{D}\approx K_{D}$. Thus, in the operator state representation, we can also prove $G_{D}\approx K_{D}$.

\subsection{Quantum Gram-Schmidt Process Circuits}
We apply the quantum Gram-Schmidt process to orthogonalize the operator states $ |g_i) $. The first orthogonalized operator state is chosen as $ |k_0) = |g_0) $, which can be a simple initial state. The subsequent orthogonalized states are obtained using the recurrence relation  
\begin{equation}
    |k_m) = \frac{|g_m) - \sum_{i = 0}^{m - 1} (g_m | k_i) |k_i)}{\| |g_m) - \sum_{i = 0}^{m - 1} (g_m | k_i) |k_i) \|_2},
\end{equation}
where $ |k_m) $ represents the orthogonalized operator state. Since generating $ |k_m) $ requires knowledge of the previously orthogonalized states $ |k_i) $, but no additional quantum memory is available to store them, we apply quantum state tomography. The state $ |k_i) $ is expressed in terms of $ |g_i) $ as $ |k_i) = \sum_{j = 1}^{m} w_j |g_i) $, where the coefficients $ w_j $ can be determined using the Hadamard test, requiring up to $ m^2 $ Hadamard tests ~\cite{meng2020improved}. This allows us to reconstruct $ |k_i) $ when needed. To facilitate this process, we define the unitary operator  
\begin{equation}
  U_i = I_2 \otimes U_{k_i} (I_2 \otimes I_{{2^N}} - |0\rangle\langle 0| + X \otimes |0\rangle\langle 0|) I_2 \otimes U_{k_i}^\dagger.  
\end{equation}
Here, $ U_{k_i} $ is the unitary operation used to prepare the operator state $ |k_i) $. By applying the sequence of unitaries $ \prod_{j = 1}^{n -1 } U_j $ to the state $ |0) |g_m) $, we obtain  
\begin{equation}
    |\phi_m) = \prod_{j = 1}^{m -1 } U_j |0) |g_m) = |0) |k_m) - \sum_{j = 1}^{m - 1} (k_j | g_m) |k_j) + |\phi_m^{\bot}),
\end{equation}
where $ |\phi_m^{\bot}) $ is an unnormalized state with a nonzero component in the first register. The term $ k_m \propto |g_m) - \sum_{j = 1}^{n - 1} (k_j | g_m) |k_j) $ represents the orthogonalization step. Therefore, we can obtain the set of vectors $|k_m)$ in the complexity $\mathcal{O}(m^2\sqrt{D}poly(log(N)))$~\cite{meng2020improved}.

As mentioned earlier, the Krylov complexity $\overline{C}$, averaged over the saturated stage, scales in the same way as the Krylov dimension. Therefore, to measure the long-time average complexity, we first run the quantum Gram-Schmidt process to determine the Krylov dimension. However, to measure the Krylov complexity itself, we must take an additional step: measuring the overlap between the measured state and the Krylov bases. This overlap can be measured using the Swap test.  

For the free model, the Krylov dimension scales polynomially with system size, such as in the case of free fermions where $ D \propto N^2 $. The number $ m $ of coefficients of $ |k_i) $ represented in the basis $ |g_i) $ is at most $ D $. Thus, the entire algorithm requires polynomial resources for the free case. In contrast, for the interacting case, the Krylov dimension grows exponentially with system size. However, during short-time dynamics, the initial local operator does not spread across the entire Hilbert space but only involves local operators. As a result, the number of Krylov bases remains polynomial in system size. We believe this algorithm can be implemented on near-term quantum setups, such as superconducting qubits, trapped ions and Rydberg atoms, at least for small system sizes.

\end{widetext}
\end{document}